\documentclass[paper]{JHEP3}
\pdfoutput=1
\usepackage{epsfig}
\usepackage{amsmath}
\usepackage{amssymb}
\usepackage{amsbsy}
\usepackage{bbm}
\usepackage{enumerate}
\usepackage{url}
\usepackage{xspace}

\newlength{\wfig}
\newlength{\hfig}
\newlength{\hfigs}
\def\beq{\begin{equation}}
\def\beqn{\begin{eqnarray}}
\def\eeq{\end{equation}}
\def\eeqn{\end{eqnarray}}

\newcommand\HERWIG{{\tt HERWIG}}

\newcommand\HERWIGSIX{{\tt HERWIG6}}

\newcommand\PYTHIA{{\tt PYTHIA}}
\newcommand\PYTHIASIX{{\tt PYTHIA}}
\newcommand\PYTHIAEIGHT{{\tt Pythia8}}

\newcommand\FASTJET{{\tt Fastjet}}

\newcommand\CPP{{\tt c++}}
\newcommand\NLOJETPP{{\tt NLOJET++}}
\newcommand\NJET{{\tt NJET}}
\newcommand\GOSAM{{\tt GoSam}}
\newcommand\HELACNLO{{\tt HELAC-NLO}}
\newcommand\KRA{K_{\scriptscriptstyle \rm R}}
\newcommand\KFA{K_{\scriptscriptstyle \rm F}}

\newcommand\muF{\mu_{\sss\rm F}}
\newcommand\muR{\mu_{\sss\rm R}}
\newcommand\muf{\mu_{\sss\rm F}}
\newcommand\mur{\mu_{\sss\rm R}}

\newcommand\Ht{H_{\sss T}}

\def\rg{\right\}} 
\def\lg{\left\{} 
\def\({\left(} 
\def\){\right)}

\newcommand\sss{\mathchoice%
{\displaystyle}%
{\scriptstyle}%
{\scriptscriptstyle}%
{\scriptscriptstyle}%
}

\newdimen\hbigcirc
\newdimen\wbigcirc

\newdimen\figwidth

\figwidth=\textwidth

\newcommand\pt{p_{\sss \rm T}}

\newcommand\kt{k_{\sss\rm T}}

\newcommand\POWHEG{{\tt POWHEG}}
\newcommand\POWHEGBOX{{\tt POWHEG BOX}}
\newcommand\POWHEGBOXVtwo{{\tt POWHEG BOX V2}}
\newcommand\MG{{\tt MadGraph4}}
\newcommand\MADGRAPH{\MG}

\newcommand\MINLO{{\tt MiNLO}}
\newcommand\MiNLO{{\tt MiNLO}}

\newcommand\trijet{{\tt trijet}}
\newcommand\dijet{{\tt dijet}}
\newcommand\VJ{{\tt VJ}}
\newcommand\ttilde{\raise.17ex\hbox{$\scriptstyle\mathtt{\sim}$}}

\newcommand{\mathd}{\mathrm{d}}

\newcount\minutes 
\newcount\scratch 
\def\timestamp{%
\scratch=\time 
\divide\scratch by 60 
\edef\hours{\the\scratch} 
\multiply\scratch by 60 
\minutes=\time 
\advance\minutes by -\scratch 
---$\,$\hours:\null 
\ifnum\minutes< 10 0\fi 
\the\minutes}

\preprint{}

\title{{Three-jet production in POWHEG}}

\author{Adam Kardos and Paolo Nason\\
  INFN, Sezione di Milano Bicocca, Piazza della Scienza 3, 20126 Milan, Italy\\
  E-mail: \email{adam.kardos@mib.infn.it, paolo.nason@mib.infn.it}
}
\author{Carlo Oleari\\
  Universit\`a di Milano-Bicocca and INFN, Sezione di Milano-Bicocca\\
  Piazza della Scienza 3, 20126 Milan, Italy\\
  E-mail: \email{carlo.oleari@mib.infn.it}}

\abstract{We present an implementation of the production of three jets at NLO
  plus parton-shower effects in the POWHEG BOX. Using the recently introduced
  \MiNLO{} procedure for setting the renormalization and factorization
  scales, we are able to obtain a generator that is also well behaved when
  the third jet becomes unresolved.  We compare key distributions computed at
  the NLO level, at the level of the \POWHEG{} hard emission and after full
  shower by \PYTHIASIX, \PYTHIAEIGHT{} and \HERWIGSIX.  We also compare our
  three-jet generator with the already available dijet \POWHEG{} generator.}

\begin{document}

\section{Introduction}
The production of hadronic jets is an ever present phenomenon in hadronic
collisions. Jets are the manifestation of the production of coloured
particles with large transverse momentum, and in hadronic collisions this
phenomenon is very frequent, due to the relatively large size of the strong
coupling constant, and to the presence of coloured incoming partons.

Electroweak processes with associated production of QCD jets are an ever
present background both to Standard Model studies and to searches of new
physics. It is therefore mandatory to understand these phenomena to our
best. For this reason, basic QCD jet-production processes can constitute a
framework where we can test our ability to simulate jet phenomena.

We stress that QCD jet production is more difficult to understand and
simulate with respect to associated jet-production phenomena. In fact, in the
latter case, our initial process already involves the production of a massive
object, with a relatively well defined mass.  This sets the relevant scale
and momentum fractions for the parton distribution functions~(pdfs), and the
associated jet production probes the values of pdfs around this point. On the
other hand, in the basic QCD jet-production processes, these scales and
momentum fractions are instead determined by the jet system, that is not as
well known.  Small errors in the determination of the jet energy induce
larger uncertainties due to the steep fall of the luminosity as a function of
the mass of the produced system. Furthermore, at relatively low total
transverse energies, we are approaching the high-energy regime, that is not
usually dealt rigorously by current Monte Carlo implementations.\footnote{For
  a shower implementation focused upon the high-energy limit,
  see~\cite{Andersen:2011hs}.}  Thus, a reasonable understanding of basic QCD
jet simulation can increase our confidence that we can also model associated
jet-production phenomena in a reliable way.

A NLO-accurate generator for dijet production that can be interfaced
to parton-shower generators (i.e.~a NLO+PS generator), using the
\POWHEG{} method~\cite{Nason:2004rx, Frixione:2007vw}, was implemented
in ref.~\cite{Alioli:2010xa} (the \dijet{} generator from now on),
within the \POWHEGBOX{} framework~\cite{Alioli:2010xd}.

In the present work, we implement a NLO+PS generator, built in the
\POWHEGBOX{} framework, for the production of three jets (the \trijet{}
generator from now on).  Basically, we include the $2\to 3$ parton scattering
processes and all the QCD corrections to them, that include, besides the
virtual corrections, all the $2\to 4$ parton scattering processes at leading
order.  We neglect parton masses throughout.  The \trijet{} implementation is
carried out within the \POWHEGBOXVtwo{} framework.\footnote{The
  \POWHEGBOXVtwo{} framework is an enhanced version of the original
  \POWHEGBOX{} package. A paper describing the new features of the
  \POWHEGBOXVtwo{} is in preparation.}

The NLO virtual matrix elements for three-jet production were computed for
the first time in refs.~\cite{Bern:1993mq, Bern:1994fz, Kunszt:1994nq}.
Compact expressions for the real contributions are also available from
refs.~\cite{Gunion:1985bp, Gunion:1986zh, Gunion:1986zg, Kuijf:1991kn}.  We
have used these results as coded in the \CPP{} program
\NLOJETPP~\cite{Nagy:2003tz}.  The other missing ingredients, needed to set
up a \POWHEGBOX{} generator, are the colour- and spin-correlated Born
amplitudes. These are easily obtained using the
\MADGRAPH{}~\cite{Alwall:2007st} interface to the \POWHEGBOX{} developed in
ref.~\cite{Campbell:2012am}.  The Born phase space, due to the complex
singularity structure of the Born amplitude, has been built using a
multi-channel technique.

The Born process in the \trijet{} generator has several singular kinematic
regions, associated to pairs of final-state partons becoming collinear, or
one parton acquiring small transverse momentum. A further, overall, singular
configuration is the one where all partons have small transverse momentum.
We find that the \MiNLO{}~\cite{Hamilton:2012np} procedure for setting the
scales and assigning Sudakov form factors is particularly helpful here, since
it tames the divergences in all kinematic regions but the overall one.  As we
will discuss in the following, we also find that, when using \MiNLO{},
two-jet inclusive observables are fairly well described, so that we do not
need to worry about the impact of configurations close to the limit where one
jet becomes unresolved, and furthermore we only need to deal with the
problems of the overall singularity of the Born-level cross section, since
the others are regulated by \MiNLO{}.

The paper is organized as follows. In section~\ref{sec:technical} we describe
the construction of the Born phase space, and the multi-channel technique
that we have used in order to probe adequately all the singular regions, and
other technical details about the \trijet{} implementation.
In section~\ref{sec:checks} we discuss the checks that we have carried
out in order to validate our cross-section formulae.
In section~\ref{sec:comparisons} we compare the output of our generator at
different levels.  After the discussion of the common settings for the
comparisons in section~\ref{sec:settings}, we compare among each other the
NLO, the Les Houches Event~(LHE) level and the shower results in
section~\ref{sec:NLOvsLH}. The LHE level is the stage where \POWHEG{} has
already generated the hardest radiation, but no other radiation has been
added by the subsequent shower programs. The purpose of this comparison is to
determine how the final result is built up.
%
In section~\ref{sec:NLOvsNLOMiNLO} we compare the NLO and the \MiNLO-improved
NLO results, in order to show at what level the \MiNLO{} procedure differs
from the NLO results obtained with a standard scale choice. In
section~\ref{sec:NLOvsLHMiNLO} we compare among each other the NLO,
LHE and shower results, when \MiNLO{} is turned on.  Since \MiNLO{} regulates the
divergences related to the third jet becoming soft or collinear, but not
those related to the whole event having small total transverse energy, we
discuss how to enforce some physicality requests on the small
transverse-energy region in section~\ref{sec:smallHT}.  In
section~\ref{sec:trijet-dijet-comparison}, we compare the \MINLO{} \trijet{}
results with the \dijet{} ones, when considering quantities inclusive in the
third jet. 
We give our conclusions in section~\ref{sec:conclusions}.  A short discussion
on the choice and setting of the dynamical scales in the \POWHEGBOX{} is
presented in appendix~\ref{app:btlscalereal}.

\section{Technical details}
\label{sec:technical}
In this section we discuss a few technical details of the \trijet{}
implementation: the multi-channel Born phase space, the generation cuts and
the production of weighted events, with and without \MINLO{}.

\subsection{Phase-space generation with multi-channel technique}
\label{sec:multichannel}
The Born cross section for the production of three partons has several sharp
peaks, in correspondence to the singular regions where soft and collinear
singularities are approached.  A standard way to integrate a many-peak
function is by using a multi-channel technique.  In the following we
illustrate how we implemented it on trijet production.

We label the particles as follows: with 1 and 2 we indicate the two incoming
partons, using the label 0 when we refer to both, and with 3, 4, 5 the
final-state ones.  Momentum conservation in the center-of-mass frame at the
Born level is then given by
\begin{equation}
p_1 + p_2 = p_3 + p_4 + p_5\,.
\end{equation}
The Born cross section has 9 singular regions, according to the 9 possible
choices of emitter-emitted couples. We label them with two indexes: the first
index identifies the emitting particle and the second the emitted one:
\begin{itemize}
\item[-] 6 final-state regions: $\{35,53,45,54,34,43\}$, where, for example,
  35 is the singular region associated with parton 5 being emitted by parton
  3.

\item[-]  3 initial-state regions: $\{03,04,05\}$, where we treat as one the
  singular region associated with either of the incoming partons.

\end{itemize}
In order to perform an efficient importance sampling in each singular region,
we introduce a function of the kinematic variables that approaches 1 only in
one singular region and goes to zero fast enough in all the others. To do so,
we define the following quantities
\begin{equation}
S_{ij} = \frac{1}{d_{ij}}\,,  
\end{equation} 
where, for final-state partons,
\begin{equation}
 d_{ij}=  2\, p_i\cdot p_j \,\frac{E_i E_j}{E_i^2+E_j^2}= 2\,\frac{E_i^2
   E_j^2}{E_i^2+E_j^2}\( 1-\cos\theta_{ij}\) \,,   \qquad  i,j\ge 3\,, 
\end{equation}
with $E_i=p_i^0$, and $\theta_{ij}$ the angle between parton $i$ and
parton $j$, in the center-of-mass frame, and
\begin{equation}
 d_{0j} = E_j^2 \(1-y^2\), \qquad y= 1- \frac{p_1\cdot p_j}{E_1 E_j}=\cos
 \theta_{1j}\,,  
  \qquad  j\ge 3\,, 
\end{equation}
where $\theta_{1j}$ is the angle between the direction of the first incoming
beam and the outgoing parton $j$.  It is clear from their definition that
when two final-state partons become collinear or when one parton becomes
soft, the corresponding $S_{ij}$ diverges. The same can be said about
$S_{0j}$ when the $j$th parton becomes soft or collinear with respect to the
incoming beam.

Defining
\begin{equation}
S= S_{03}+S_{04}+S_{05}+S_{35} +S_{53}+S_{45}+S_{54}+S_{34}+S_{43}\,,
\end{equation}
we can then write the following identity
\begin{eqnarray}
\label{eq:ident}
1 &=& \frac{S_{03}}{S} +  \frac{S_{04}}{S} +  \frac{S_{05}}{S}  
+\frac{S_{35}}{S} \,\frac{E_5}{E_3+E_5}+\frac{S_{53}}{S} \,\frac{E_3}{E_3+E_5}+
\frac{S_{45}}{S} \,\frac{E_5}{E_4+E_5}+\frac{S_{54}}{S} \,\frac{E_4}{E_4+E_5}
\nonumber\\
&&{}+\frac{S_{34}}{S} \,\frac{E_4}{E_3+E_4} +\frac{S_{43}}{S} \,\frac{E_3}{E_3+E_4}\,,
\\[2mm]
\label{eq:ident_short}
&\equiv& \sum_j \tilde{S}_{0j} + \sum_{ij} \tilde{S}_{ij}\,,  
\end{eqnarray}
where we have introduced a self-explanatory notation in
eq.~(\ref{eq:ident_short}).  Each term on the right-hand-side of
eqs.~(\ref{eq:ident}) or~(\ref{eq:ident_short}) approaches 1 only in one
particular singular region. For example, the terms $\tilde{S}_{0j}$ approach 1
when the $j$th parton is either soft or collinear to any of the two incoming
beams, and go to 0 when other singular regions are approached. Similarly,
terms of the form $\tilde{S}_{ij}$ approach 1 when the $j$th parton is either
soft or collinear to the final-state parton $i$, and go to 0 when other
singular regions are approached. 

We insert then 1 written as eq.~(\ref{eq:ident_short}) in the formula of the
invariant phase-space element 
\begin{equation}
\mathd \Phi_B   = \mathd \Phi_B \lg  \sum_j \tilde{S}_{0j} + \sum_{ij}
\tilde{S}_{ij} \rg =  \sum_j \tilde{S}_{0j} \, \mathd \Phi_B + \sum_{ij}
\tilde{S}_{ij}  \, \mathd \Phi_B \,.
\end{equation}
We can now choose the best parametrization of the kinematic variables
(i.e. the momenta $p_i$) in terms of the Monte Carlo integration variables,
in order to do an importance sampling for each of the terms of the sum. Each
$\mathd \Phi_B$ in the sum has then a different parametrization in terms of
the Monte Carlo integration variables, so that each of them can be seen to
depend on the summation indexes. We will indicate this by adding the
subscript $ij$ to each phase-space element volume
\begin{equation}
\label{eq:multichannel}
\mathd \Phi_B = \sum_{kj} \tilde{S}_{kj} \,  \(\mathd \Phi_B\)_{kj}\,.
\end{equation}
In the \trijet{} generator, each phase-space volume $\(\mathd \Phi_B\)_{kj}$ is
computed using a replica of the automatic machinery (see refs.~\cite{Frixione:2007vw,
  Alioli:2010xd}) for the generation of the real phase space, starting from
the Born phase space for dijet production, i.e.~starting from the $2\to 2$
phase space and attaching an extra parton, with different importance sampling
according to the singular region where it has been adapted to.

When the subroutine for the computation of the Born phase space is invoked,
one extra random number is used to choose, with equal probability, one
of the 9 different parametrization of eq.~(\ref{eq:multichannel}).  The
returned Jacobian is then the product of $\(\mathd \Phi_B\)_{kj}$ and of the
corresponding suppression function $\tilde{S}_{kj}$, multiplied by 9, in
order to compensate for the 1/9 factor introduced by choosing to evaluate
randomly only one single term of the sum.

The real phase space for trijet production is then built as usual
by the \POWHEGBOX{} automatic machinery on top of the Born kinematics.

\subsection{Generation cuts or weighted events}
As already stated, the Born cross section for three-jet production has
several singular regions, associated to a pair of final-state partons
becoming collinear among each other, or to a final-state parton becoming
collinear to an initial-state parton, or becoming soft. Because of these
singularities, an unweighted generator would end up generating all events in
these singular regions.  This problem is usually handled by requiring that
the final-state partons satisfy some generation cuts, such as to avoid the
singular configurations.  In this case, of course, one should make sure that
the contributions arising from the neglected regions of phase space do not
end up affecting observables of interest. In the case at hand, one may
require that the three partons form well separated jets. If final-state
observables do require at least three jets, it is unlikely that the neglected
regions would contribute to them.  However, one should always check that the
results are independent upon these cuts.

An alternative method is to generate weighted events. One chooses a weight
function that diverges when approaching the singular regions. This is done as
follows. One introduces a function of the Born phase-space kinematics,
$F(\Phi_B)$, that vanishes in the singular regions, such that the following
integral of the differential cross section $\sigma$
\begin{equation}
\int \frac{\mathd \sigma}{\mathd \Phi_B} F(\Phi_B) \;\mathd \Phi_B
\end{equation}
is finite.  One then generates the phase-space points with a probability
proportional to $(\mathd \sigma/\mathd \Phi_B) F(\Phi_B)$, assigning to each
point a weight $1/F(\Phi_B)$.  In this way one should not worry about the
independence of the result upon generation cuts. Observables that do depend
upon the singular regions will typically receive rare contributions with
large weights, yielding large errors.  This method was used in the
\POWHEGBOX{} since ref.~\cite{Alioli:2010qp}, where the function $F$ was
dubbed ``Born suppression factor''. In the \POWHEGBOX{} framework the factor
$F$ was applied to all cross-section contributions (i.e.~not only the Born
term), and, in the case of real terms and collinear remnants, it was computed
as a function of the underlying-Born kinematics.  Here we stress that, in
spite of the name, no phase-space regions are really suppressed. In fact, the
effect of the $F$ factor at the generation level is exactly compensated by
the $1/F$ weight of the event.

In the present case, we have considered the following
$F$ function
\begin{eqnarray}
\label{eq:F}
F&=&F_1\times F_2\,, \\
F_1 &=& \exp\left[-S_1^p \times \left(\frac{1}{q_1^p}+\frac{1}{q_2^p}+\frac{1}{q_3^p}+
\frac{1}{q_{12}^p}+\frac{1}{q_{23}^p}+\frac{1}{q_{13}^p}\right)\right], \\
\label{eq:F2}
F_2 &=& \frac{\frac{1}{S_2^p}}{\left(\frac{1}{S_2}+\frac{1}{\Ht^2}\right)^p}\,,
\end{eqnarray}
where $S_1$ and $S_2$ are suitably chosen scales, $q_1$, $q_2$ and $q_3$ are
the square of the transverse momenta of the three final-state particles with
respect to the beam axis, and $q_{ij}$ is the relative transverse momentum
squared of particle $i$ and $j$, defined in the partonic center-of-mass frame
as
\begin{equation}
q_{ij}=p_i\cdot p_j \, \frac{E_i E_j}{E_i^2+E_j^2}\,.
\end{equation}
Furthermore
\begin{equation}
\Ht = \sqrt{q_1}+\sqrt{q_2}+\sqrt{q_3}\,.
\end{equation}
The role of $F_2$ is to handle the singular region associated to all partons
having small transverse momentum. It also plays the role of increasing the
importance sampling in the region of large transverse-momentum jets, a
feature that is needed in order to properly cover a large range of transverse
energy.

\subsection{MiNLO}
If we apply the \MiNLO{} procedure to the \trijet{} generator, the factor
$F_1$ discussed above is no longer needed. All singular regions, except for
the overall one, are regulated by the \MiNLO{} Sudakov form factors. The
\MiNLO{} form factor is exactly as described in ref.~\cite{Hamilton:2012np},
with the only freedom of choosing the scale of the basic process,\footnote{In
  the \MiNLO{} framework, by basic process we mean the process before any
  branching has occurred, i.e.~$H$ production in $H$ + jets, and dijet
  production in the present case.}  that in the case of
ref.~\cite{Hamilton:2012np} (dealing with Higgs boson production in
association with jets) was taken equal to the Higgs boson virtuality. 
In the present case, we have chosen the scale of the basic process to be
equal to the sum of the transverse momenta of the two final pseudoparticles
(after the \MiNLO{} clustering has taken place).

\section{Checks of the code}
\label{sec:checks}
We have performed several checks on our code.  We have generated subroutines
for the virtual corrections using \NJET~\cite{Badger:2012pg},
\GOSAM~\cite{Cullen:2011ac, Mastrolia:2010nb, Mastrolia:2012du,
  Binoth:2008uq, Heinrich:2010ax, Guillet:2013msa, vanHameren:2010cp} and
\HELACNLO~\cite{Bevilacqua:2011xh} and compared them to the virtual
contributions obtained with the routines in our program.  We have found that
the \NJET{}, \GOSAM{} and \HELACNLO{} results were in agreement among each
other for all subprocesses. We also found agreement with the \NLOJETPP{}
routines, except for the $ q\bar{q} g g g$ amplitude, and all its crossings,
where there was a problem in the colour sum. After fixing this, we found
perfect agreement.

The Born contribution, together with the colour- and spin-correlated
Born amplitude, were generated using the \MADGRAPH{} \POWHEGBOX{}
interface. This allowed also the generation of the real contribution
according to \MADGRAPH{}, that was thus checked against the (much
faster) one that we have implemented in our code.

As a further check, the full NLO calculation was compared with the one of
ref.~\cite{Badger:2012pf} by first comparing the plots displayed there with
the results of our code, that we ran using the same parton distribution
functions and the same scales. We found agreement within statistical errors.
In order to carry out this comparison we needed to run the \POWHEGBOX{} with
the same dynamical scale choice of ref.~\cite{Badger:2012pf}. Although the
default scale choice in the \POWHEGBOX{} is computed as a function of the
underlying-Born kinematics, it is possible to set it up in such a way that
any scale choice can be implemented. See appendix~\ref{app:btlscalereal} for
more details.

\section{Comparing  \trijet{} results at various levels}
\label{sec:comparisons}
The \trijet{} generator can be used to compute three-jet observables at the
NLO level, at the \POWHEG{} Les Houches Event level (i.e.~with the hardest
emission generated according to the shower technique implemented in
\POWHEG{}), and after the full shower.  The \MiNLO{} feature can be turned on
already at the NLO stage.  In the present section we compare the output of
our generator at these levels.

\subsection{General settings for the forthcoming comparisons}
\label{sec:settings}
We consider jet production at the 8~TeV LHC. We use the {\tt CT10nlo} pdf
set~\cite{Lai:2010vv}.  We remind the reader that any of the modern pdf sets
can be used~\cite{Martin:2009iq, Ball:2012cx}, and our choice is only a
matter of definiteness. We consider the shower output at the parton level,
without the inclusion of hadronization effects and multiple interactions,
since, at this stage, we are not comparing our result with jet
data.\footnote{A comparison with available data is in progress.}  We
interface our generator to \PYTHIASIX{}~(version 6.4.25),
\PYTHIAEIGHT{}~(version 8.183) and \HERWIG{}~(version 6.510).  When using
\PYTHIASIX{} we use the Perugia~0 tune~\cite{Skands:2010ak}~({\tt
  pytune(320)}).  We use \PYTHIAEIGHT{} and \HERWIG{} with their default
tunes.

In the \POWHEG{} settings we have included the {\tt doublefsr 1}
option, and the modification of the {\tt scalup} prescription obtained by
setting {\tt changescalup 1}. Both these features are
illustrated in ref.~\cite{Nason:2013uba}.

We encountered severe stability problems when using \HERWIG{}, showing up as
spikes in our final histograms. We investigated them, and found that they
were related to events having very small transverse energy (of the order of
few GeV) at the Les Houches level, and developing very high transverse
momentum jets (above 50~GeV) after shower.  The cause of these problems was
photon emission from quarks, that apparently does not comply with the {\tt
  scalup} veto in \HERWIG{}. These problems disappeared completely by setting
{\tt vpcut=1D30}, that switches off photon radiation.  Thus, all our
\HERWIG{} results were obtained with this setting. We verified that it has no
visible effect on our results, but for the disappearance of the spikes.

The use of \PYTHIASIX{} with the \trijet{} generator in combination with
\MINLO{} requires particular care.  In fact, in this case, \PYTHIASIX{} is
unable to shower a sensible fraction of very small $\Ht$ events. These events
are not going to contribute to physically interesting distributions. Thus,
they should be treated as events that did not pass the cuts, and should be
counted when dividing the total weight entering a histogram bin by the total
number of events. It turns out that, when \PYTHIASIX{} finds such events, it
silently discards them and loads a new one. In the analysis setup of the
\POWHEGBOX, the total number of events is usually computed as the number of
times that the analysis routine is called, and, because of the aforementioned
behaviour of \PYTHIASIX, the fraction of discarded events is thus not
counted. We coded a workaround for this problem in our analysis routines. A
user adopting different analysis frameworks must make sure not to incur this
problem.

We adopt the following values for the parameters entering the $F$
function in eqs.~(\ref{eq:F}) and~(\ref{eq:F2}):
\begin{equation}
S_1=\left(50\;{\rm GeV}\right)^2\,,\quad\quad S_2=\left(800\;{\rm
  GeV}\right)^2\,,\quad\quad p=2\,.
\end{equation}
In addition, in order to avoid uninteresting regions with very small
transverse momenta that may cause numerical problems, we reject all Born
configurations such that $\min\{q_i,q_{ij}\}<\left(0.3\;{\rm GeV}\right)^2$.
The value of the factorization and renormalization scale is taken equal to
$\Ht/2$, computed on the partonic configuration of the underlying-Born
kinematics.

Finally, jets are reconstructed using the anti-$\kt$
algorithm~\cite{Cacciari:2008gp} as implemented in the \FASTJET{}
package~\cite{Cacciari:2011ma, Cacciari:2005hq}, with jet radius $R$.

The results we will show in the next sections have been obtained by
generating 2.4~M events in two runs, with and without \MiNLO{}.  The runs
have been performed on a 48 core machine, and they took roughly 37 and 100
hours, respectively.

\subsection{NLO, LHE and shower-level comparisons}
\label{sec:NLOvsLH}

\begin{figure}
\epsfig{file=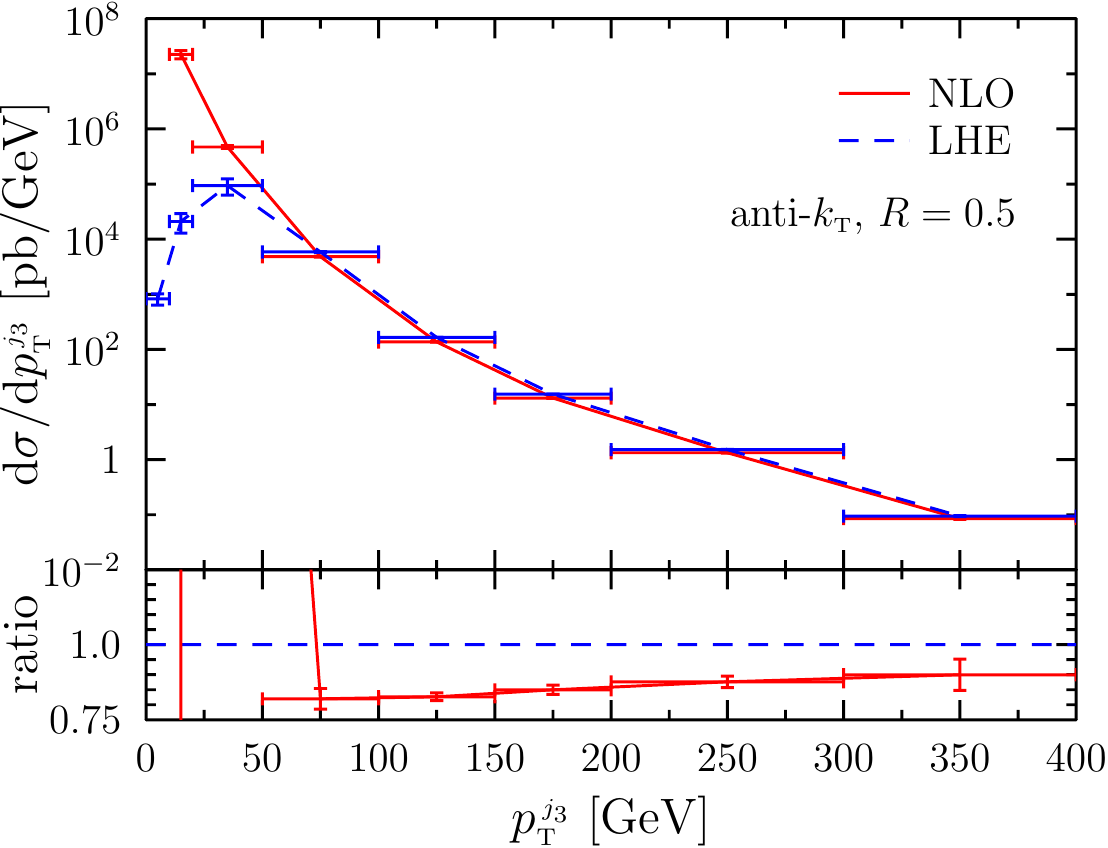,width=0.49\textwidth}
\epsfig{file=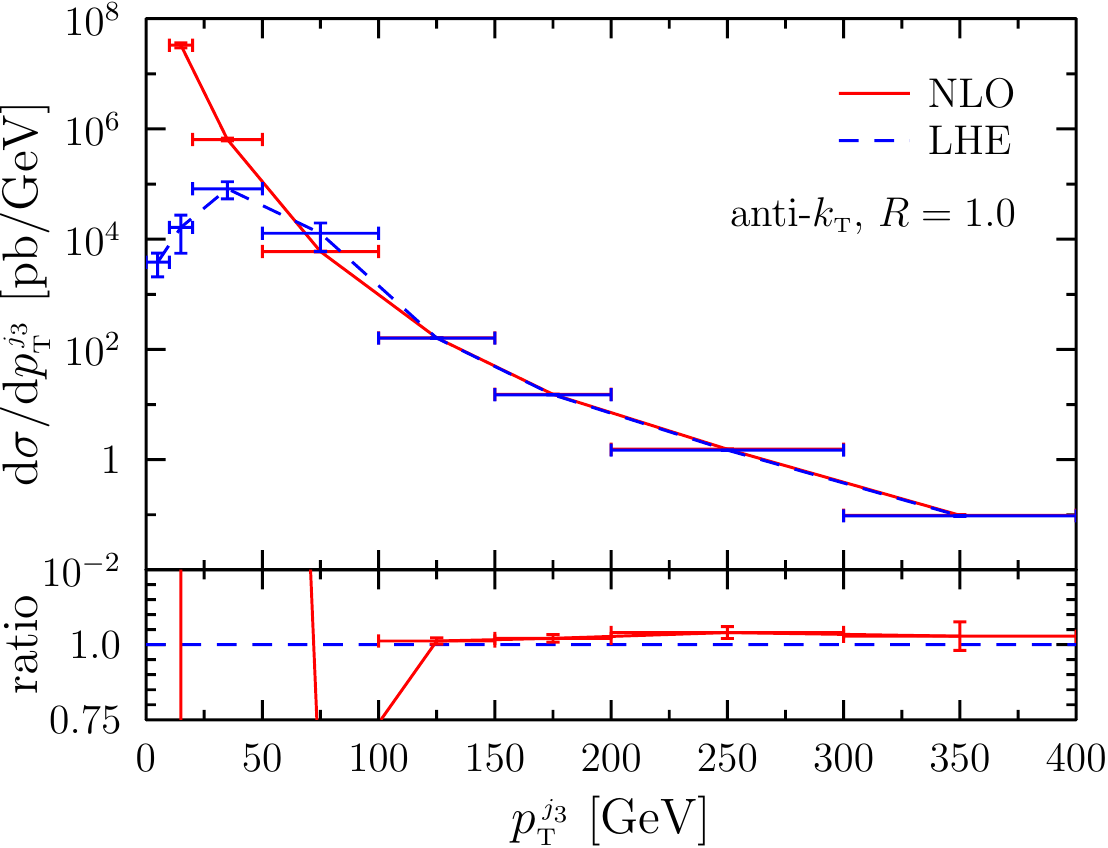,width=0.49\textwidth}
\caption{\label{fig:NLO-LH-pt3} Comparison of the NLO and LHE results for the
  transverse-momentum distribution of the third jet, for $R=0.5$~(left) and
  $R=1$~(right).}
\end{figure}
We begin by showing in fig.~\ref{fig:NLO-LH-pt3} the comparison of the
fixed-NLO and the LHE-level results for the transverse-momentum distribution
of the third jet.
\begin{figure}
\epsfig{file=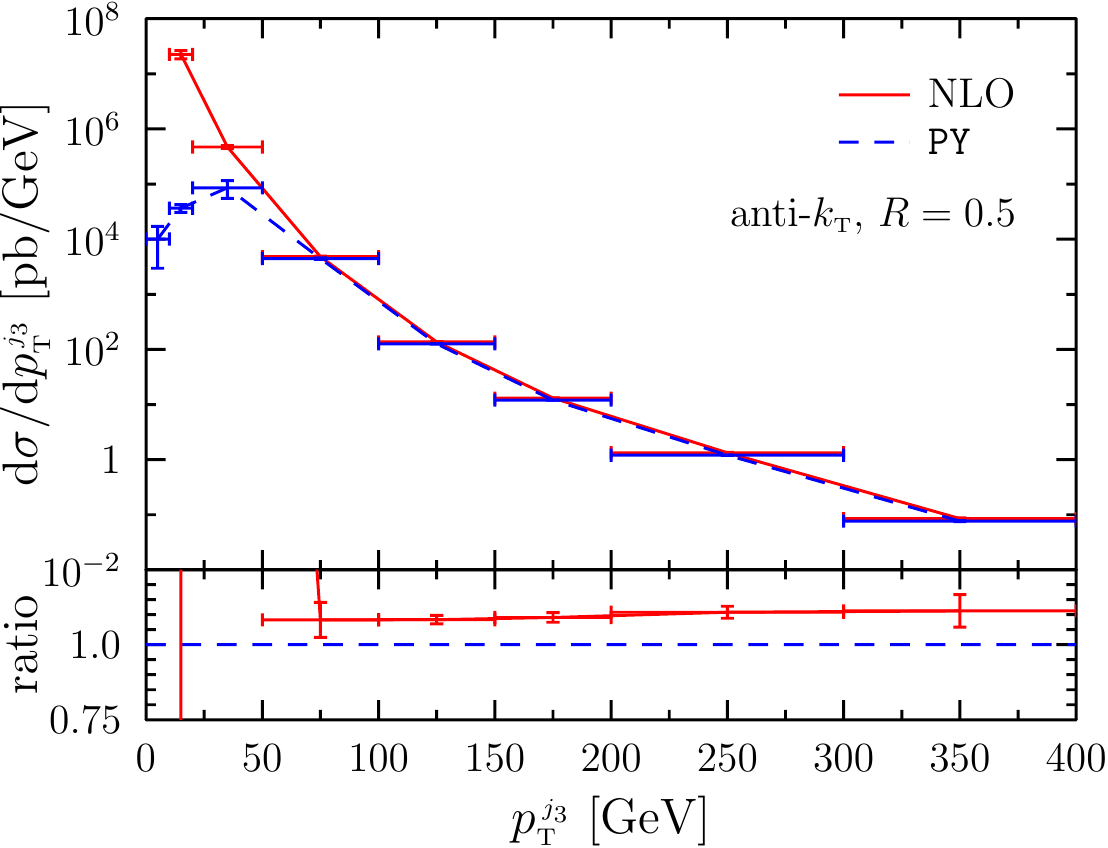,width=0.49\textwidth}
\epsfig{file=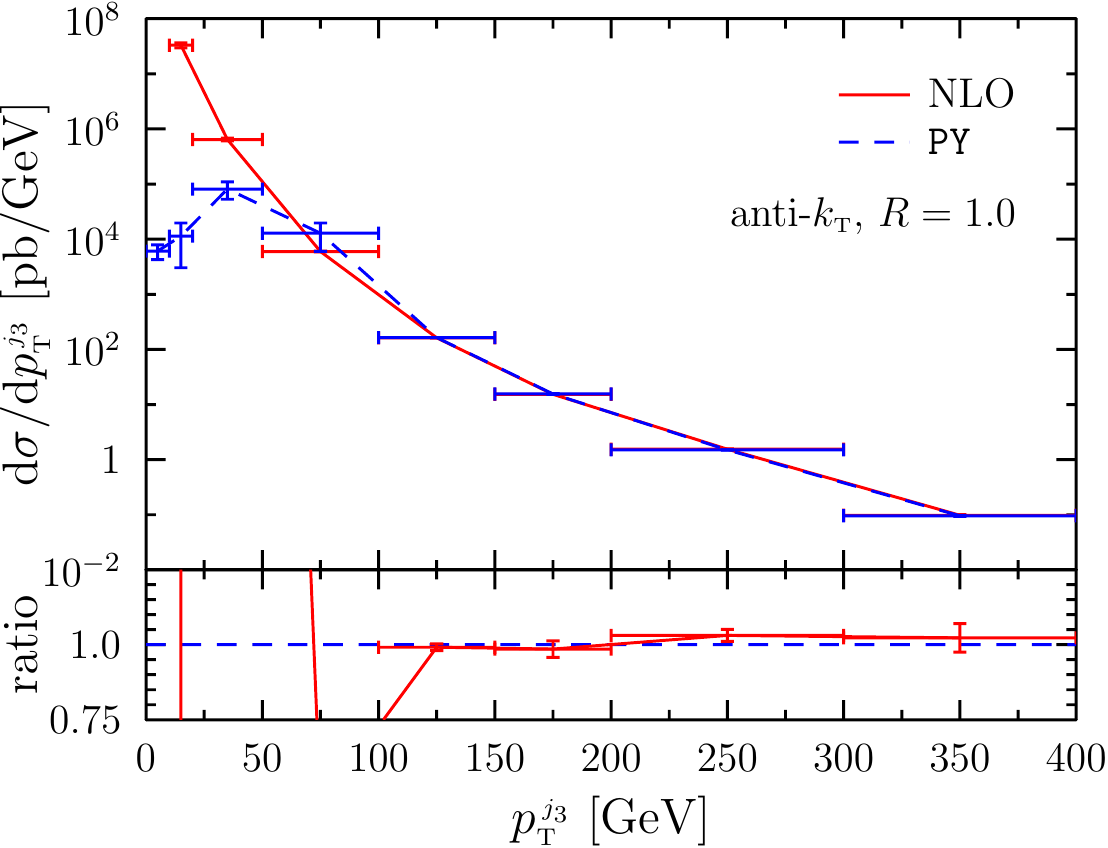,width=0.49\textwidth}
\caption{\label{fig:NLO-PY-pt3} Comparison of the NLO and \PYTHIA{} showered
  results for the transverse-momentum distribution of the third jet, for
  $R=0.5$~(left) and $R=1$~(right).}
\end{figure}
\begin{figure}
\epsfig{file=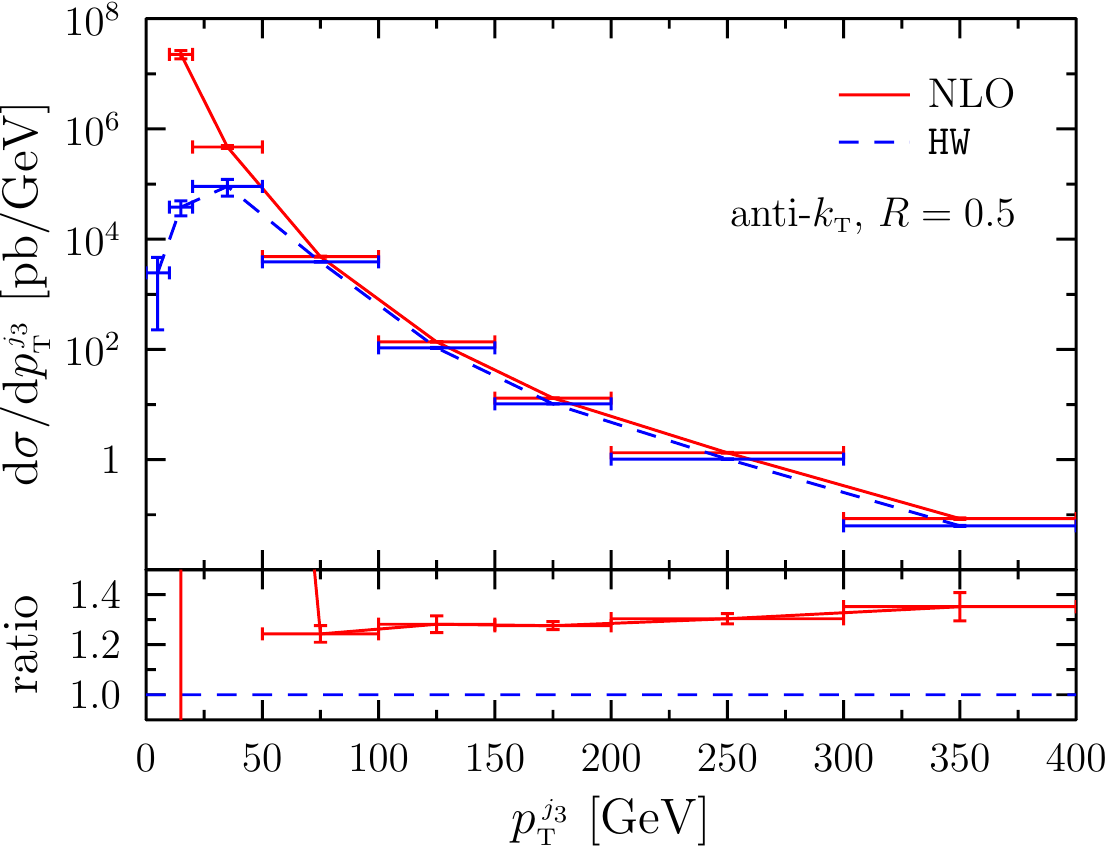,width=0.49\textwidth}
\epsfig{file=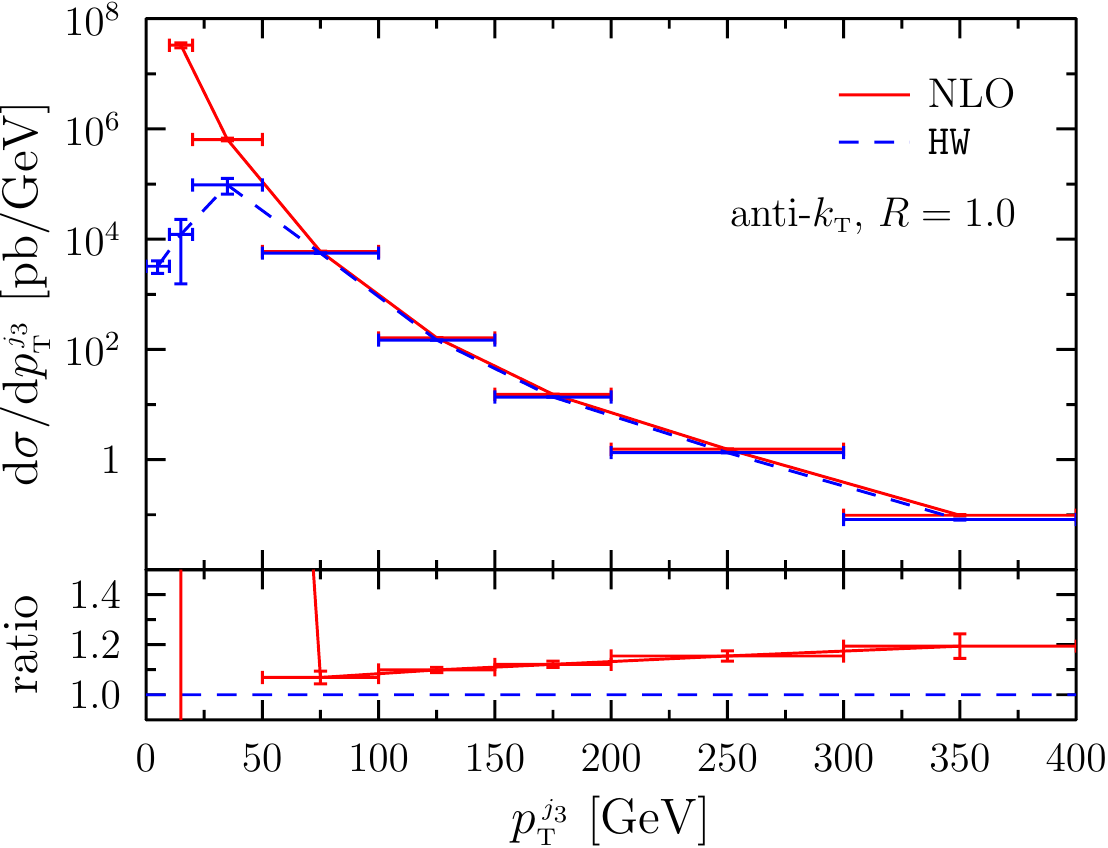,width=0.49\textwidth}
\caption{\label{fig:NLO-HW-pt3} Comparison of the NLO and \HERWIG{} showered
  results for the transverse-momentum distribution of the third jet, for
  $R=0.5$~(left) and $R=1$~(right).}
\end{figure}
The \PYTHIA{} and \HERWIG{} showered output, compared to the NLO result, are
displayed in figs.~\ref{fig:NLO-PY-pt3} and \ref{fig:NLO-HW-pt3}.

We first remark that the small transverse-momentum suppression of the LHE,
\PYTHIA{} and \HERWIG{} showered results is simply due to the fact that,
because of our $F$ function, events with transverse momentum smaller than
50~GeV are very rarely generated.

We observe that when $R=1$ all results are in better agreement. Differences
arise for smaller values of $R$, and can be understood as follows.  First of
all, it can be easily checked that the LHE level result has very mild
dependence upon $R$.  This is due to the fact that the splitting of the third
parton into two collinear partons has a very strong Sudakov suppression. In
fact, in this case the \POWHEG{} Sudakov form factor is the product of the
form factors for vetoing harder final-state splittings of all final-state
partons, times the form factor for vetoing harder initial-state radiation.
Because of this suppression, partons are relatively well separated, and a
small $R$ dependence is observed.\footnote{This feature of the LHE level
  events was already noticed and discussed in ref.~\cite{Alioli:2010xa}.}
When completing the shower, further splitting processes can take place at a
suitable rate, and the $R$ dependence is reinstated. Notice that no Sudakov
suppression for radiation is included in the fixed-order calculation,
yielding a visible $R$ dependence. It is clear, however, that in the
\trijet{} generator the $R$ dependence will mostly arise at the shower
stage. This is a desirable feature. In fact we do not expect the NLO result
to be reliable in this region, since, among other things, it lacks the
Sudakov form factor and the appropriate scale choice in the coupling
constant. Because of this, the shower algorithm will acquire the
responsibility to reliably describe the $R$ dependence.

In figs.~\ref{fig:NLO-PY-pt3} and \ref{fig:NLO-HW-pt3} we observe a
disturbing difference between \PYTHIA{} and \HERWIG{}. In the latter, the $R$
dependence is much stronger.  Our expectation is that the shower result
should be determined by two elements: on one side, the introduction of the
correct Sudakov form factor for the splitting process (that will tend to
reduce collinear splitting and thus increase the third jet cross section at
smaller $R$), and multiple emissions, that will tend to increase collinear
splitting processes, and thus reduce the jet cross section at smaller
$R$. The net effect is an increase of the shower cross section at $R=0.5$
(with respect to the $R=1$ value) that is around 10\%{} in \PYTHIA{}, but is
more of the order of 20\%{} in \HERWIG{}. Furthermore, the \HERWIG{} result
shows an increasing discrepancy with the fixed order result at large
transverse momenta of the third jet in the $R=1$ case. We have no good
understanding of why this is the case. On the other hand, the \PYTHIAEIGHT{}
result is compatible with the \PYTHIA{} one.

\begin{figure}
\epsfig{file=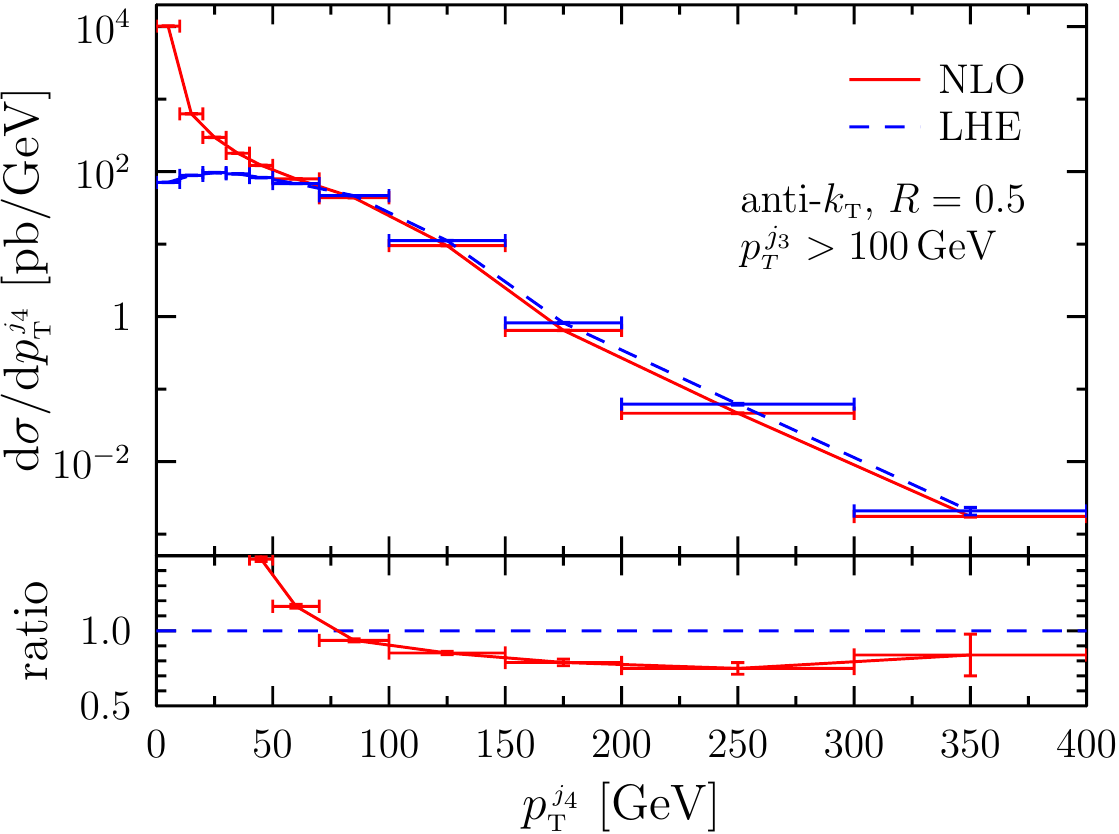,width=0.49\textwidth}
\epsfig{file=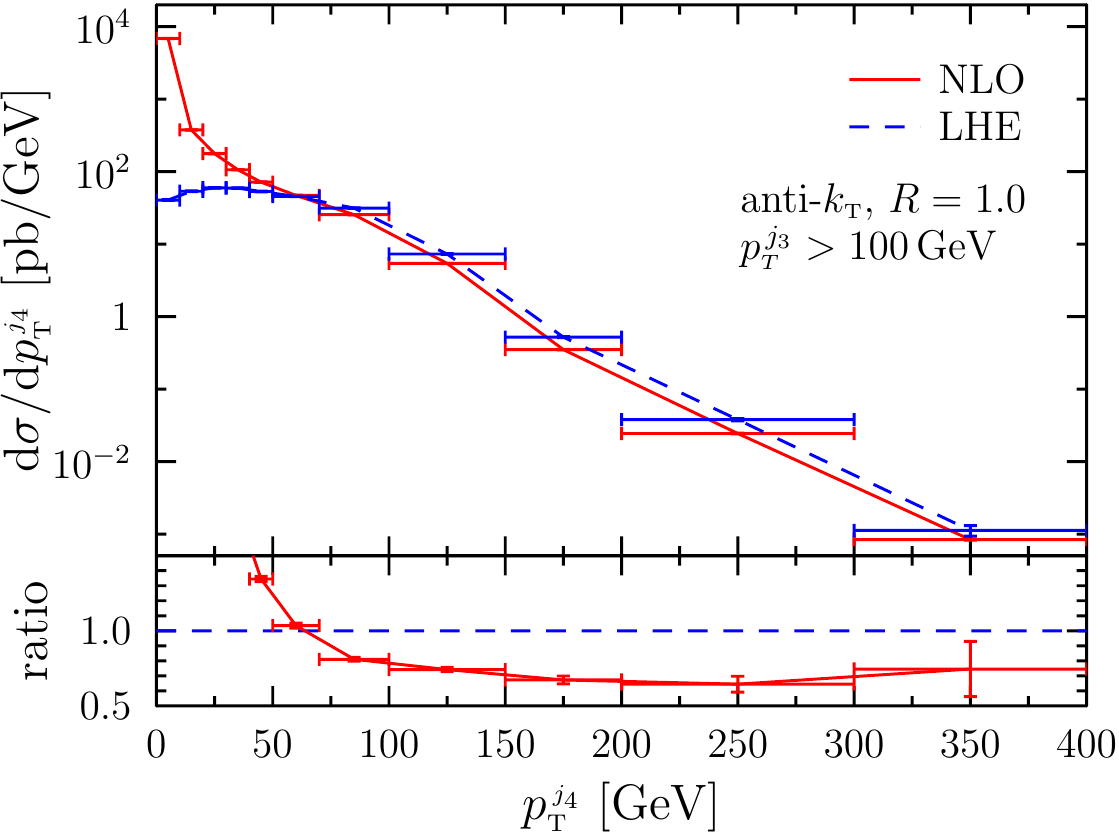,width=0.49\textwidth}
\caption{\label{fig:fourthNLO-LH} Comparison of the NLO and LHE results for
  the transverse-momentum distribution of the fourth jet, with the
  cut $\pt^{j_3}\ge 100$~GeV, for $R=0.5$~(left) and $R=1$~(right).}
\end{figure}
\begin{figure}
\epsfig{file=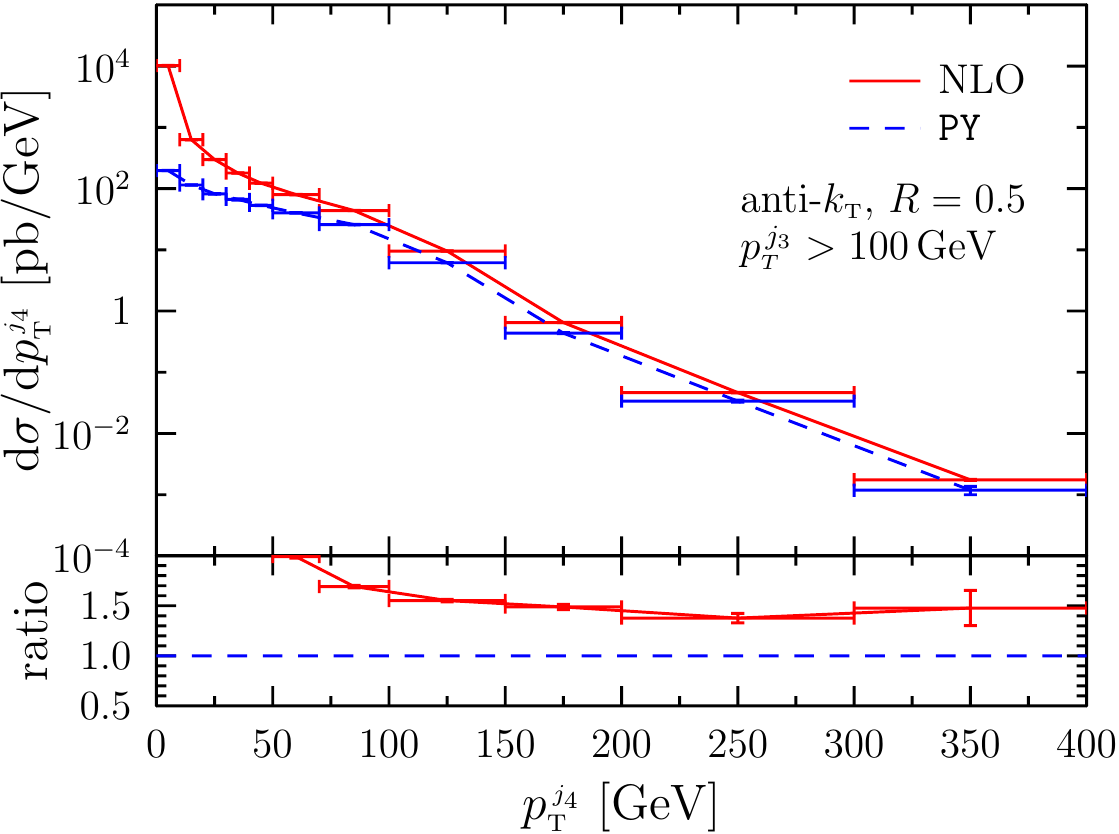,width=0.49\textwidth}
\epsfig{file=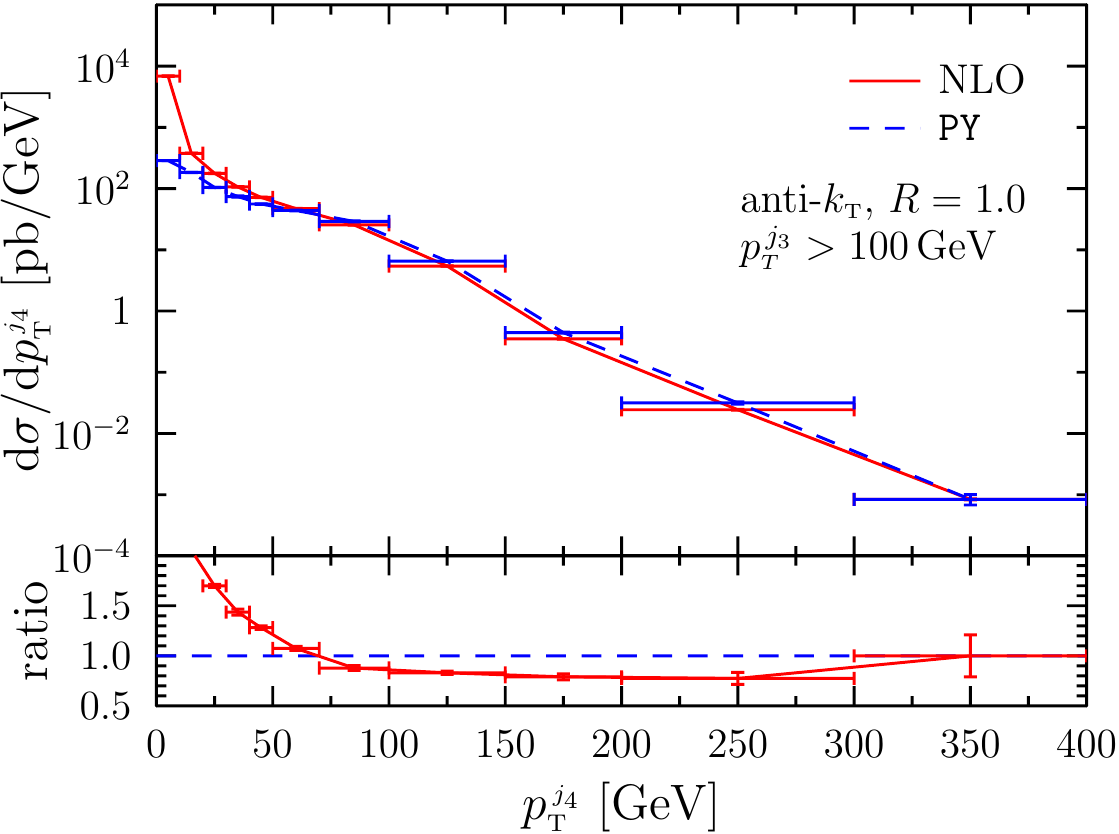,width=0.49\textwidth}
\caption{\label{fig:fourthNLO-PY} As in fig.~\protect{\ref{fig:fourthNLO-LH}},
  comparing NLO and \PYTHIA{} showered results.}
\end{figure}
When going from a fixed-NLO result to an LHE one, the most striking
differences are usually observed in the spectrum of the extra parton emitted
in the real process, that, in our case, corresponds to the fourth jet. We
thus compare the transverse-momentum distribution of the fourth jet, with a
cut on the third jet, $\pt^{j_3}\ge 100$~GeV, computed at the NLO level, with
the LHE level result (fig.~\ref{fig:fourthNLO-LH}), and with the \PYTHIA{}
showered one (fig.~\ref{fig:fourthNLO-PY}). The 100~GeV cut on the third jet
is imposed for the following reason.  If no cuts on the remaining jets are
imposed, as $\pt^{j_4}$ increases, also $\pt^{j_1}$, $\pt^{j_2}$ and
$\pt^{j_3}$ must increase, and we are thus probing an overall property of the
cross section. With the cut on the third jet, by studying the $\pt^{j_4}$
spectrum below the third-jet cut, we are studying the soft-collinear
radiation dynamics from the three-jet Born configuration. We thus expect, for
example, that the fixed-order result (that for this quantity has only leading
order accuracy) diverges at very small transverse momenta, in contrast with
the LHE result, where the soft-collinear region for the emission of the
fourth jet is strongly Sudakov suppressed.  We also remark that, in this
case, the $R$ dependence of the NLO result is not at all reliable, since no
further partons are emitted in this framework beyond the fourth.

\subsection{Comparison between the NLO results with a standard choice of
  scales and NLO+\MiNLO{}}
\label{sec:NLOvsNLOMiNLO}

\begin{figure}
\epsfig{file=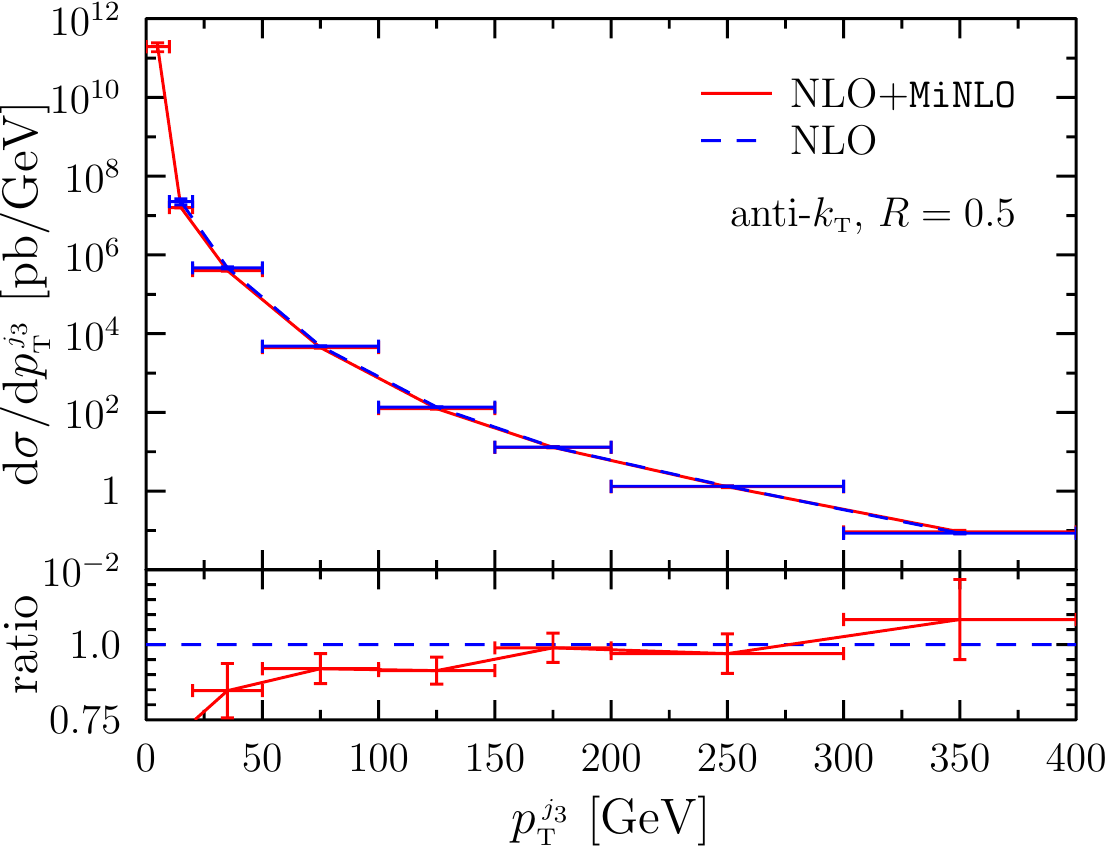,width=0.49\textwidth}
\epsfig{file=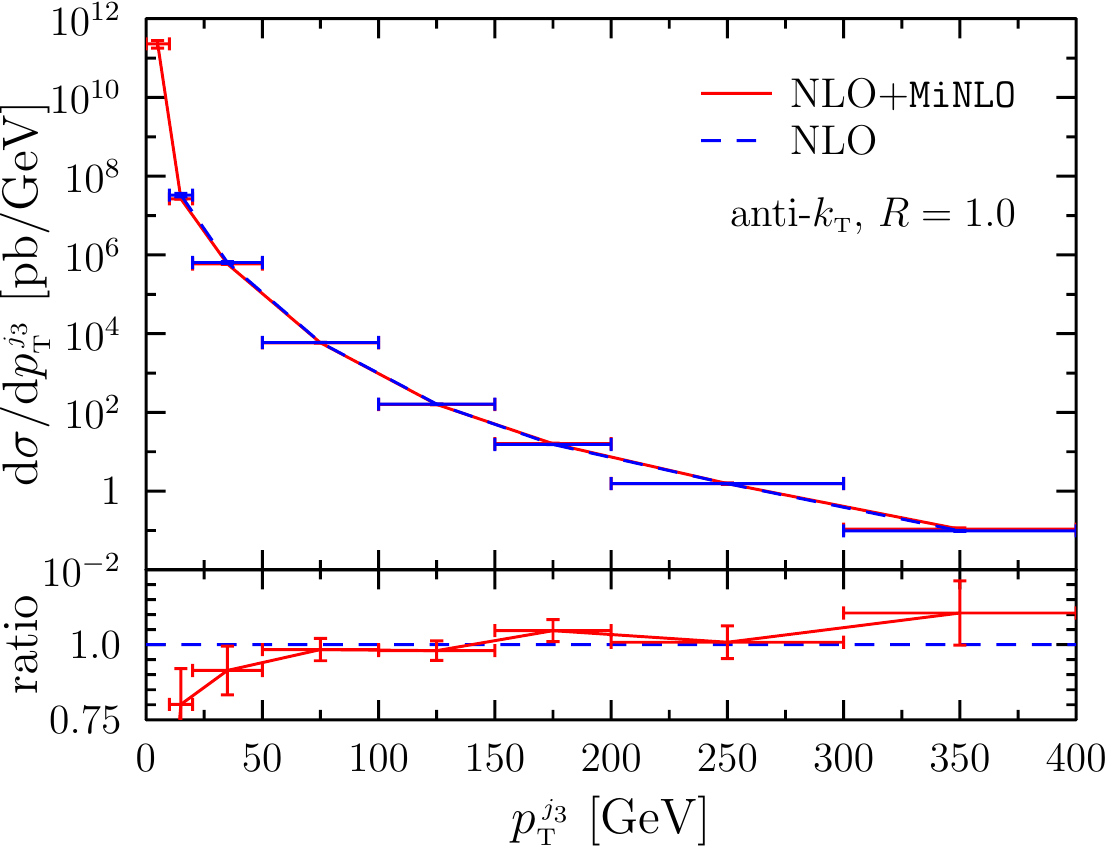,width=0.49\textwidth}
\caption{\label{fig:NLOminlo-NLO} Comparison of the NLO results with
  \MiNLO{} and the NLO result with our standard scale choice, for the
  transverse-momentum distribution of the third jet. In the left plot
  $R=0.5$, and in the right plot $R=1$.}
\end{figure}
Before turning to the discussion of the \trijet{} results when \MiNLO{} is
active, we perform a comparison of the bare NLO calculation with or without
\MiNLO{}, whose purpose is to show that only minor differences are seen in
the region where the three jets are resolved.  Here we report in
fig.~\ref{fig:NLOminlo-NLO} such comparison for the transverse-momentum
distribution of the third jet.  Our standard scale choice, when \MiNLO{} is
not used, is to set $\muR$ and $\muF$ equal to $\Ht/2$, where $\Ht$ is
computed on the kinematics of the underlying-Born configuration.  We see that
the two results are fairly compatible, except in the very small
transverse-momentum region. Here, the NLO with the standard scale choice
grows much faster in magnitude, and becomes large and negative in the first
bin. With \MiNLO{}, the small transverse-momentum region is better behaved,
as expected.  However, we remind the reader that, in
fig.~\ref{fig:NLOminlo-NLO}, this region is divergent also with \MiNLO{},
since it is dominated by the production of low transverse-momentum jets.

\subsection{NLO, LHE and shower-level comparisons with \MiNLO{}}
\label{sec:NLOvsLHMiNLO}
\begin{figure}
\epsfig{file=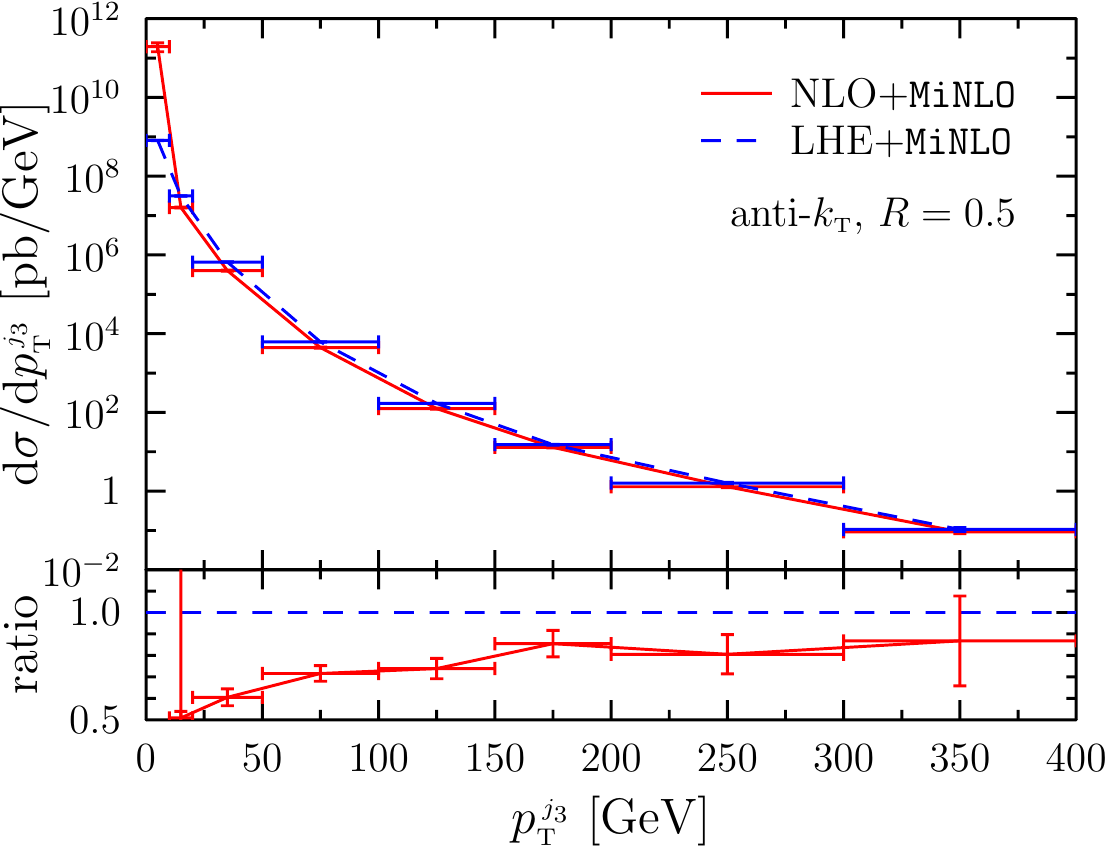,width=0.49\textwidth}
\epsfig{file=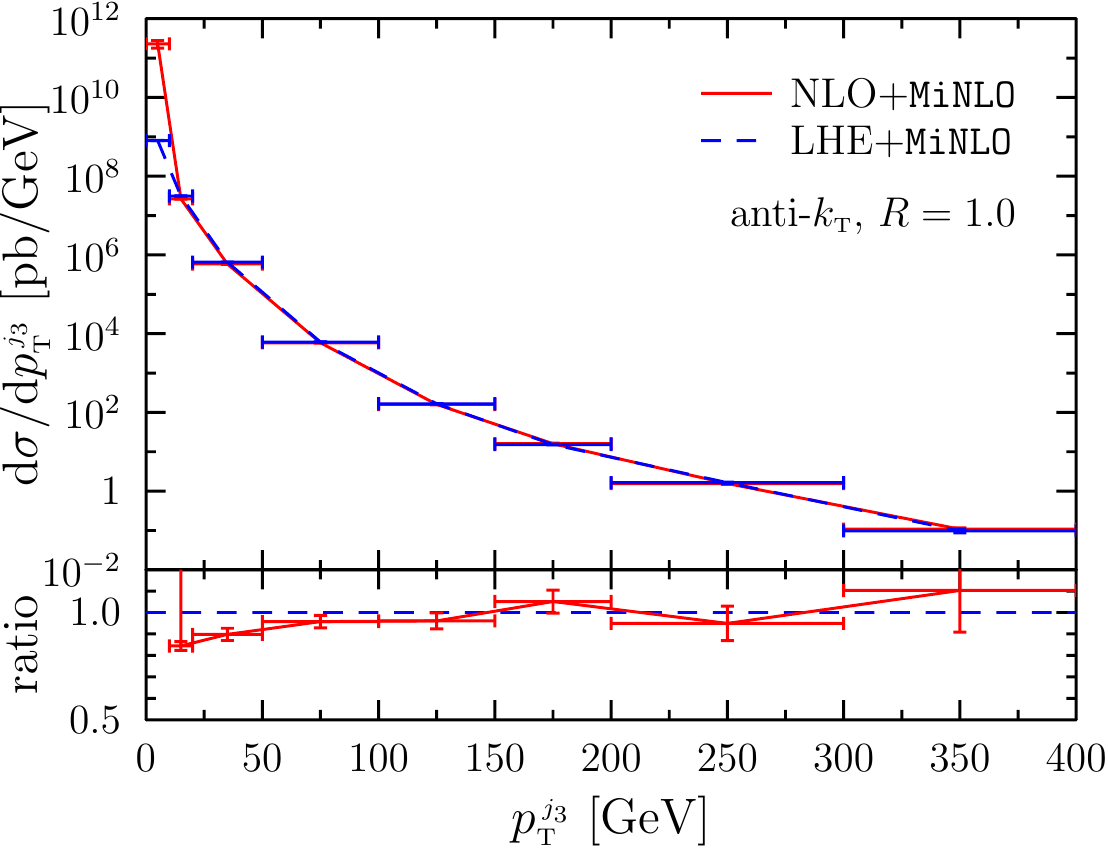,width=0.49\textwidth}
\caption{\label{fig:NLO-LH-minlo-pt3} Comparison of the NLO and LHE results
  for the transverse-momentum distribution of the third jet, for
  $R=0.5$~(left) and $R=1$~(right). \MiNLO{} is turned on.} 
\end{figure}
\begin{figure}
\epsfig{file=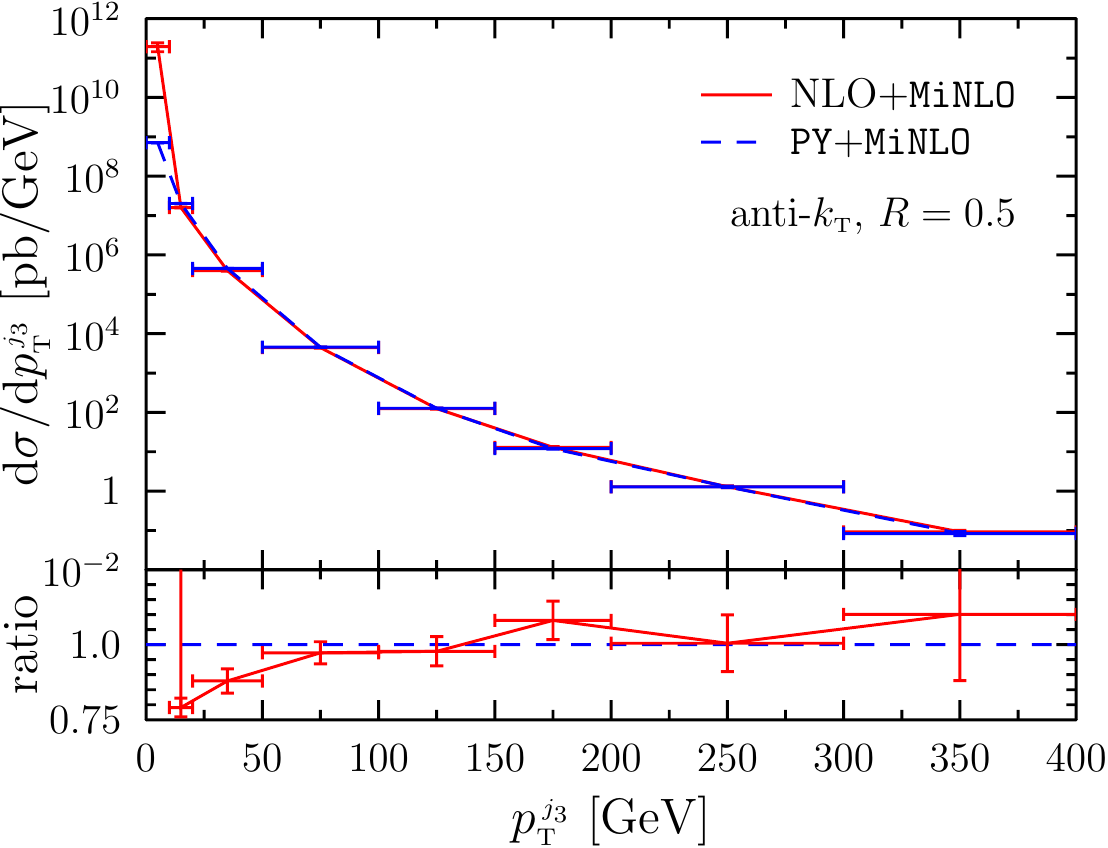,width=0.49\textwidth}
\epsfig{file=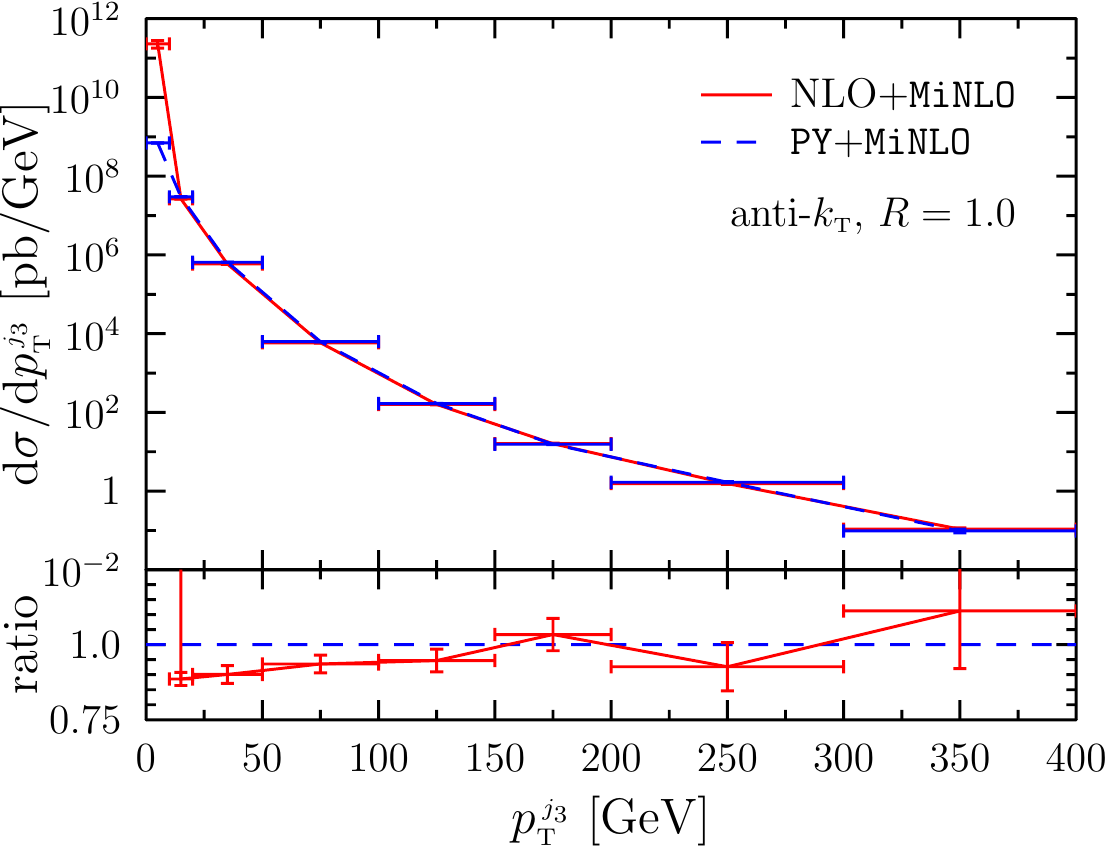,width=0.49\textwidth}
\caption{\label{fig:NLO-PY-minlo-pt3}
Comparison of the NLO and \PYTHIA{} showered
results for the transverse-momentum distribution of the third jet,
for $R=0.5$~(left) and $R=1$~(right). \MiNLO{} is turned on.}
\end{figure}
\begin{figure}
\epsfig{file=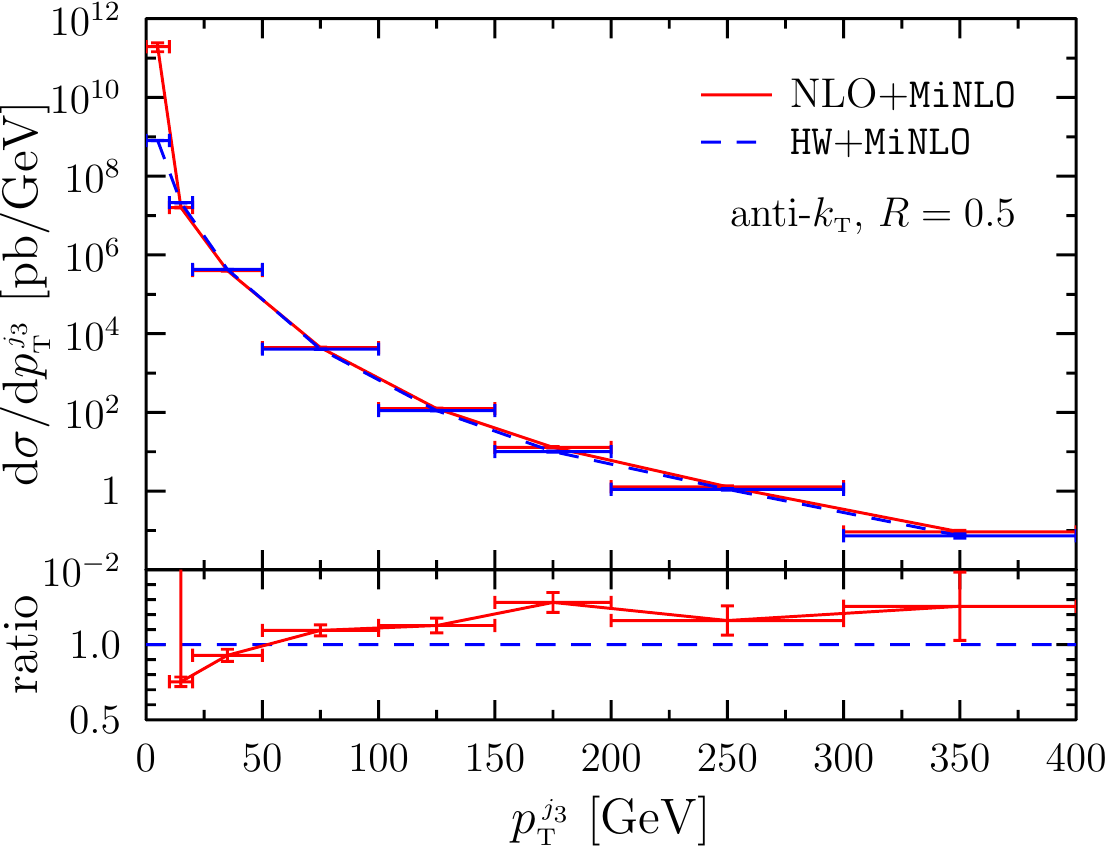,width=0.49\textwidth}
\epsfig{file=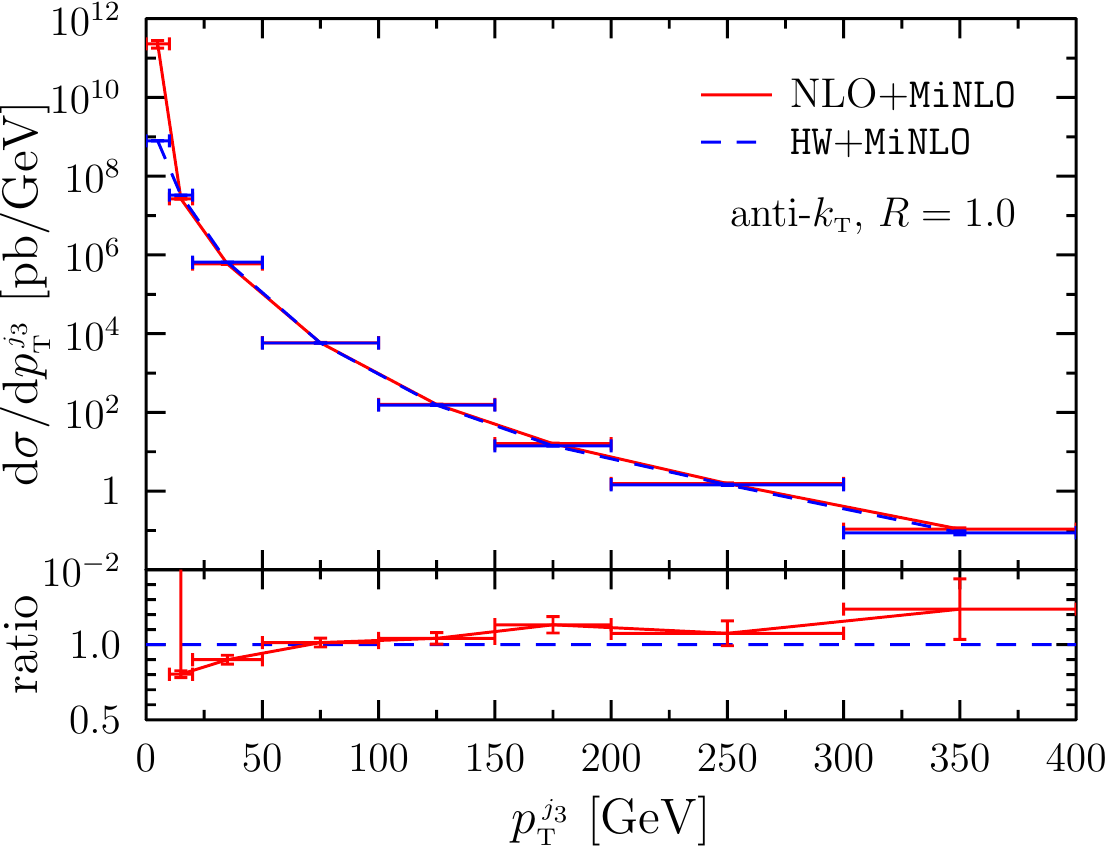,width=0.49\textwidth}
\caption{\label{fig:NLO-HW-minlo-pt3}
Comparison of the NLO and \HERWIG{} showered
results for the transverse-momentum distribution of the third jet,
for $R=0.5$ (left) and $R=1$ (right). \MiNLO{} is turned on.}
\end{figure}
It is interesting to present the results for the transverse momentum of the
third jet, when the \MiNLO{} feature is turned on.  Also in this case, for
$R=1$ we see good agreement among the NLO, LHE and the showered results.  We
illustrate these results in figs.~\ref {fig:NLO-LH-minlo-pt3},
\ref{fig:NLO-PY-minlo-pt3} and~\ref{fig:NLO-HW-minlo-pt3}.  It turns out,
however, that for $R=0.5$ the NLO result is below the LHE one by a larger
amount with respect to the no-\MiNLO{} case, which leads to a slightly better
agreement of the showered and NLO results.

\subsection{The low transverse-energy region}
\label{sec:smallHT}
When showering a \POWHEG-generated LHE configuration, we usually assume that
the jet structure of the event is only marginally affected by the shower.  In
jet production, in particular, we assume that the configurations having very
small transverse energy (i.e.~$\Ht$) at the Les Houches level should not
contribute significantly to events that pass the jet cuts. It turns out,
however, that such configurations have diverging cross section, and thus one
may worry that the very small probability that the shower has for building up
relatively hard jets starting from LHE configurations with small $\Ht$, may
end up being amplified by an unphysically large cross section.

A similar problem arises when we consider associated jets in a hard
phenomenon.  In this case, however, the \MiNLO{} procedure ensures that the
cross section for the Les Houches event is physically well behaved. On the
other hand, in the case of jet production, the Sudakov resummation is not
enough to guarantee a physical behaviour at small transverse energies.

In order to study this potential problem, we have taken a very simple
approach. We have determined the cross section for events with transverse
energy above a given cut at the Les Houches level. We have then found that
for $\Ht>10$~GeV such cross section is about 60~mb, going down to around
30~mb for $\Ht>20$~GeV. Without imposing this cut, the cross section reaches
1000~mb. The diverging behaviour is, in fact, limited by the tiny cut-off
that we impose upon the kinematics to avoid divergences.  Since the inelastic
cross section at the 7~TeV LHC is around 70~mb~\cite{Antchev:2013haa}, it
seems reasonable to impose a cut on the transverse energy of our events at
the Les Houches level, in the range between 10 and 20~GeV.

We have found that, for the shower Monte Carlo programs that we have
considered, and with our settings, the Les Houches level cut has visible
impact only on events with very small transverse momenta. For example, we
find sensible differences in the distribution of the transverse momentum of
the third jet only for $\pt^{j_3}\lesssim 5$~GeV, for the cut $\Ht>10$~GeV,
and for $\pt^{j_3}\lesssim 10$~GeV for the cut $\Ht>20$~GeV.  This is
reasonable to expect, since the $\Ht$ of the event is at least three times
the transverse momentum of the third jet.  In spite of this, we have
preferred to maintain the Les Houches level $\Ht>20$~GeV cut as our standard,
since it seems, in all cases, unreasonable to have events with a cross section
that becomes of the same order of the total inelastic cross section.

\subsection{Comparison of the \trijet{}+\MiNLO{} results with the
\dijet{} results for inclusive quantities}
\label{sec:trijet-dijet-comparison}
When using \MiNLO{}, the \trijet{} generator becomes predictive also for
inclusive quantities, i.e.~for observables that do not necessarily require
the presence of a third jet. It is hard to quantify theoretically the
accuracy of such predictions. We have shown that, for inclusive quantities,
the \VJ{} generators (i.e.~generators for Higgs or $W/Z$ boson production in
association with a jet) improved with the \MiNLO{} procedure yield an
accuracy that is better then LO, but not quite at the NLO level, unless one
makes a careful tuning of the procedure~\cite{Hamilton:2012rf}. In the case
of jets, the argument of ref.~\cite{Hamilton:2012rf} cannot be applied as is,
since the soft-singularity structure of the two-parton production process is
quite involved. Rather than trying to understand theoretically what is the
accuracy of the \trijet{}+\MiNLO{} generator for inclusive quantities, here
we simply compare its inclusive distributions to those obtained with the
NLO-accurate \dijet{} generator.

\begin{figure}
\epsfig{file=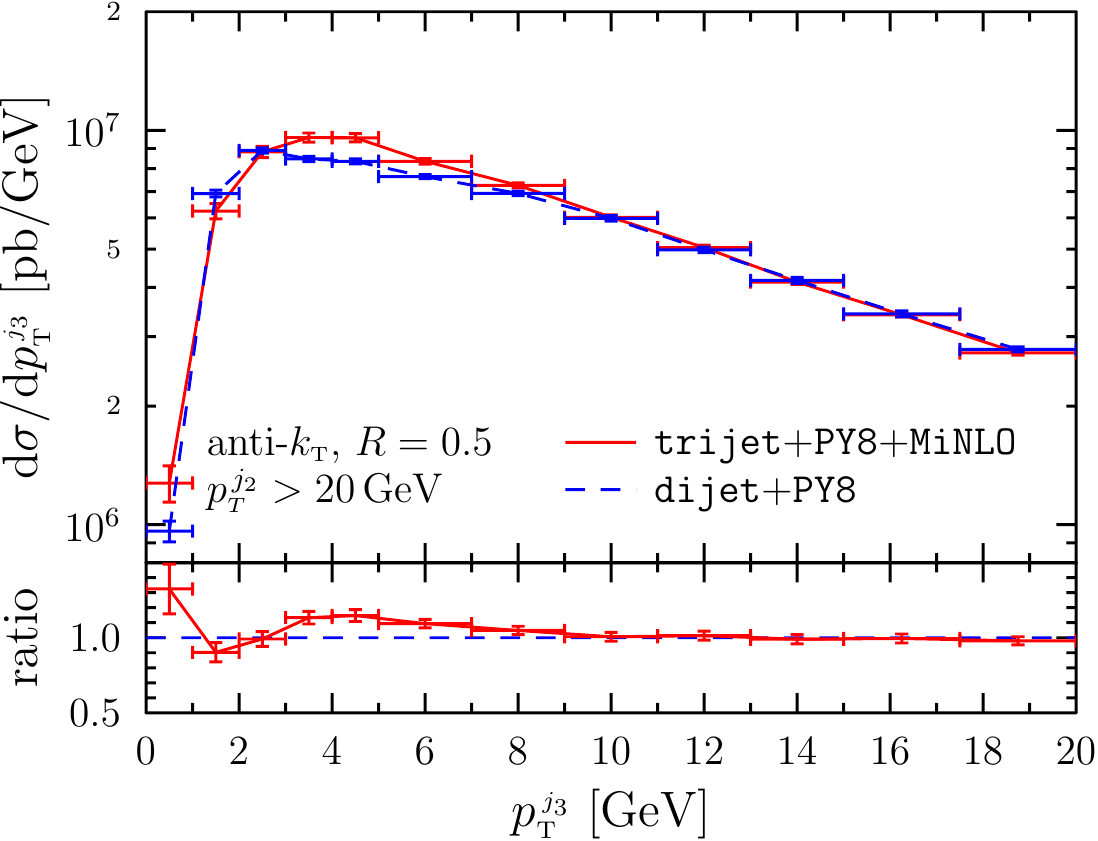,
   width=0.49\textwidth}
\epsfig{file=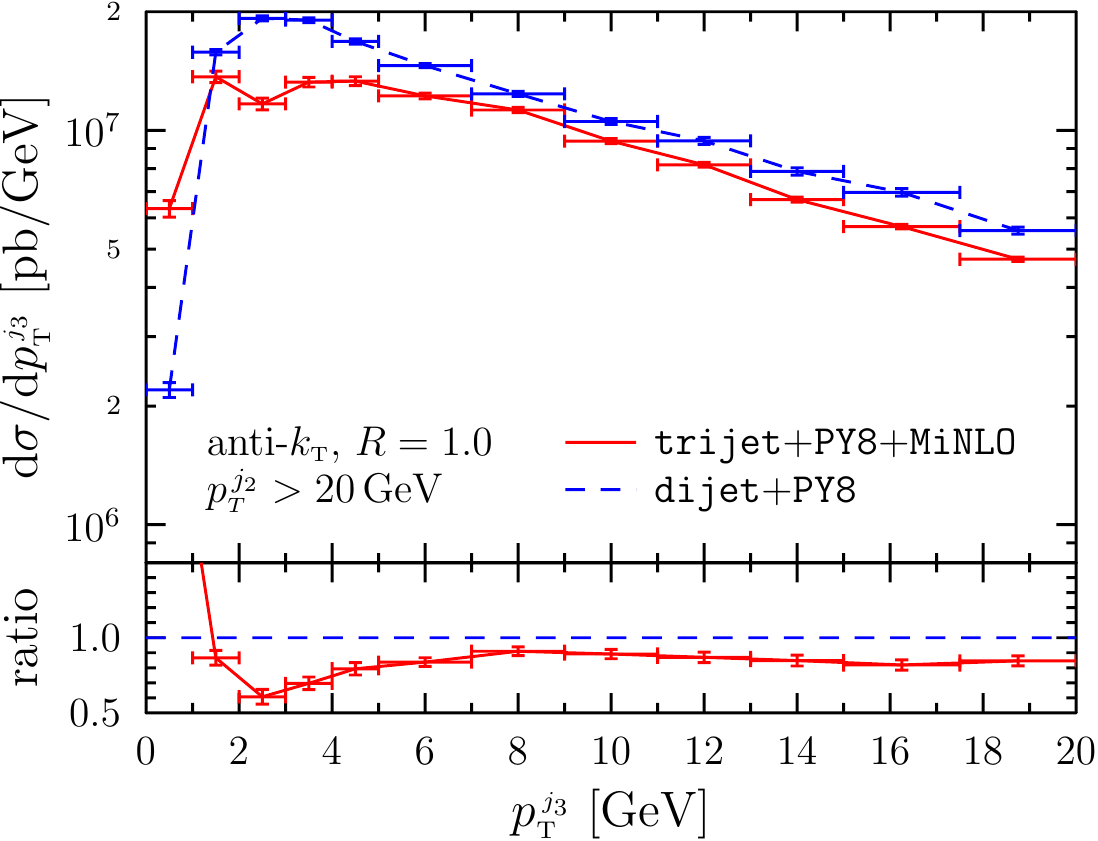,
   width=0.49\textwidth}
\caption{\label{fig:trijet-dijet-PY8-j3} Comparison between the
  \PYTHIAEIGHT{} showered results for the transverse-momentum distribution of
  the third hardest jet, for $R=0.5$~(left) and $R=1$~(right) for
  \trijet{}+\MINLO{} and \dijet{}. A minimum
  $\pt$ of 20~GeV is imposed on the second jet.}
\end{figure}
\begin{figure}
\epsfig{file=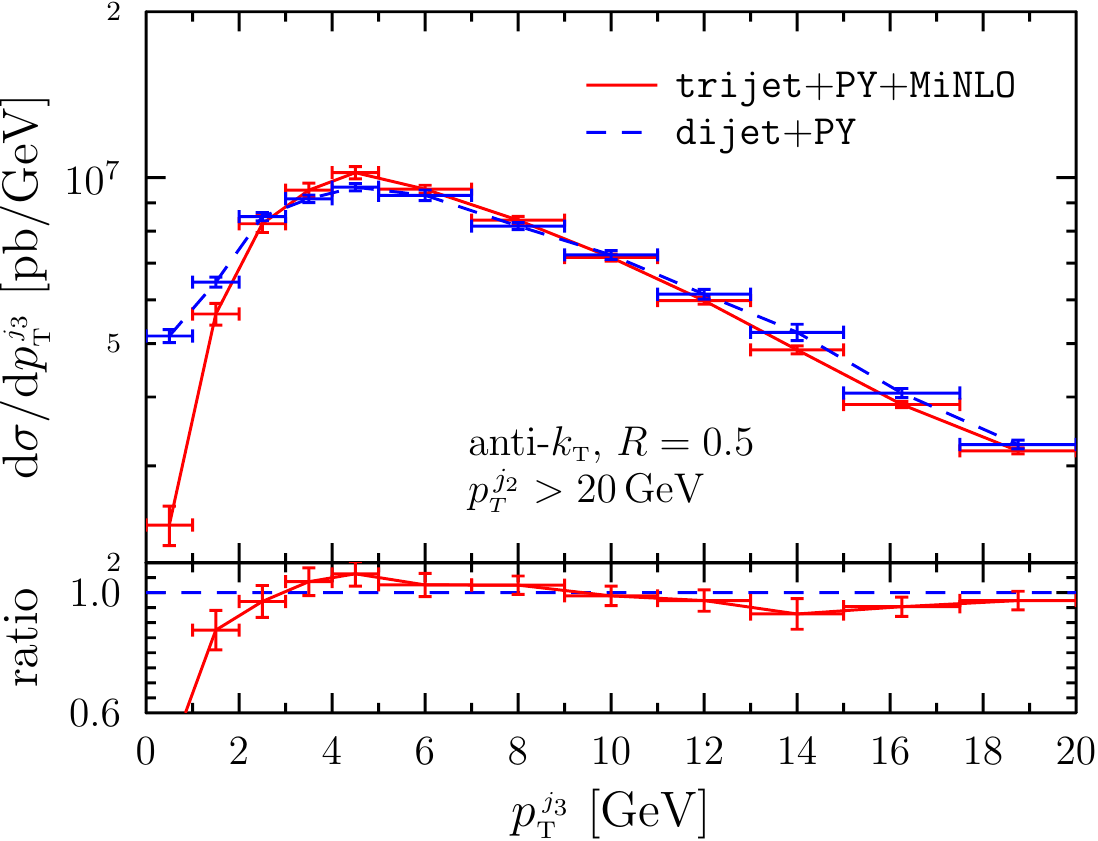,
   width=0.49\textwidth}
\epsfig{file=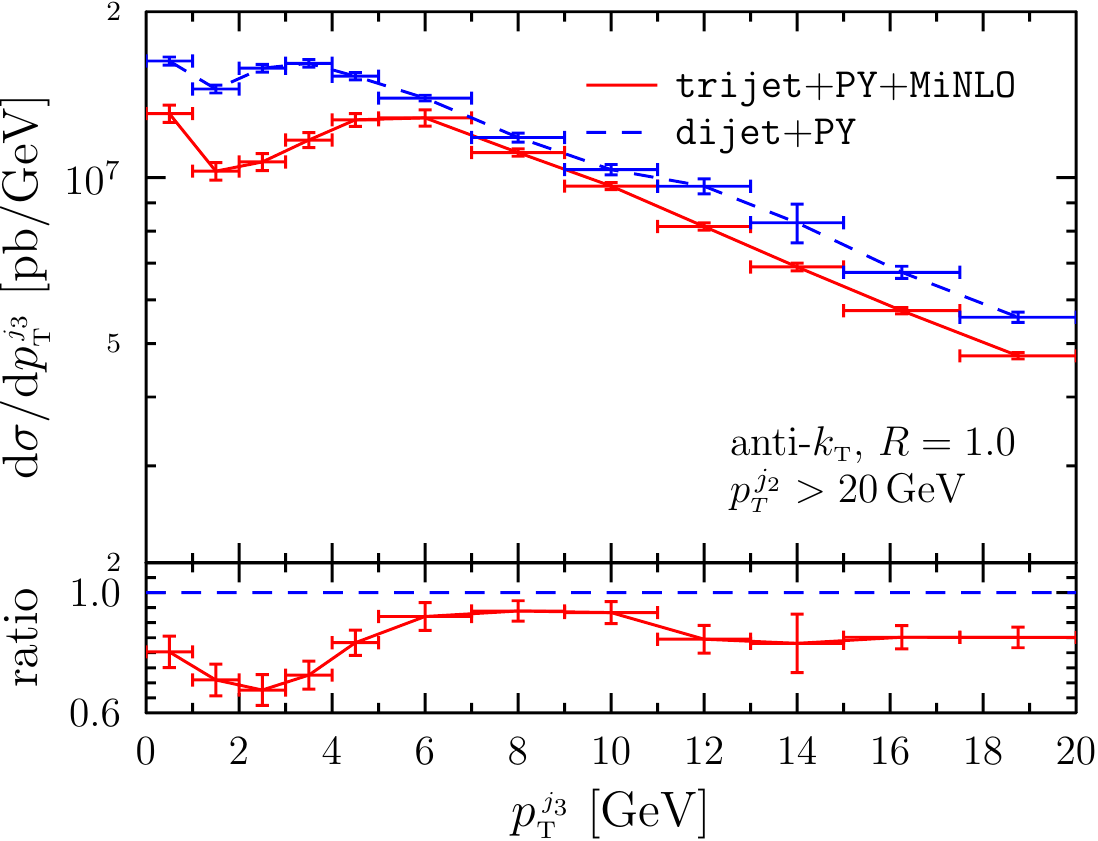,
   width=0.49\textwidth}
\caption{\label{fig:trijet-dijet-PY-j3} As in
  fig.~\protect{\ref{fig:trijet-dijet-PY8-j3}}, for \PYTHIASIX{}}
\end{figure}
\begin{figure}
\epsfig{file=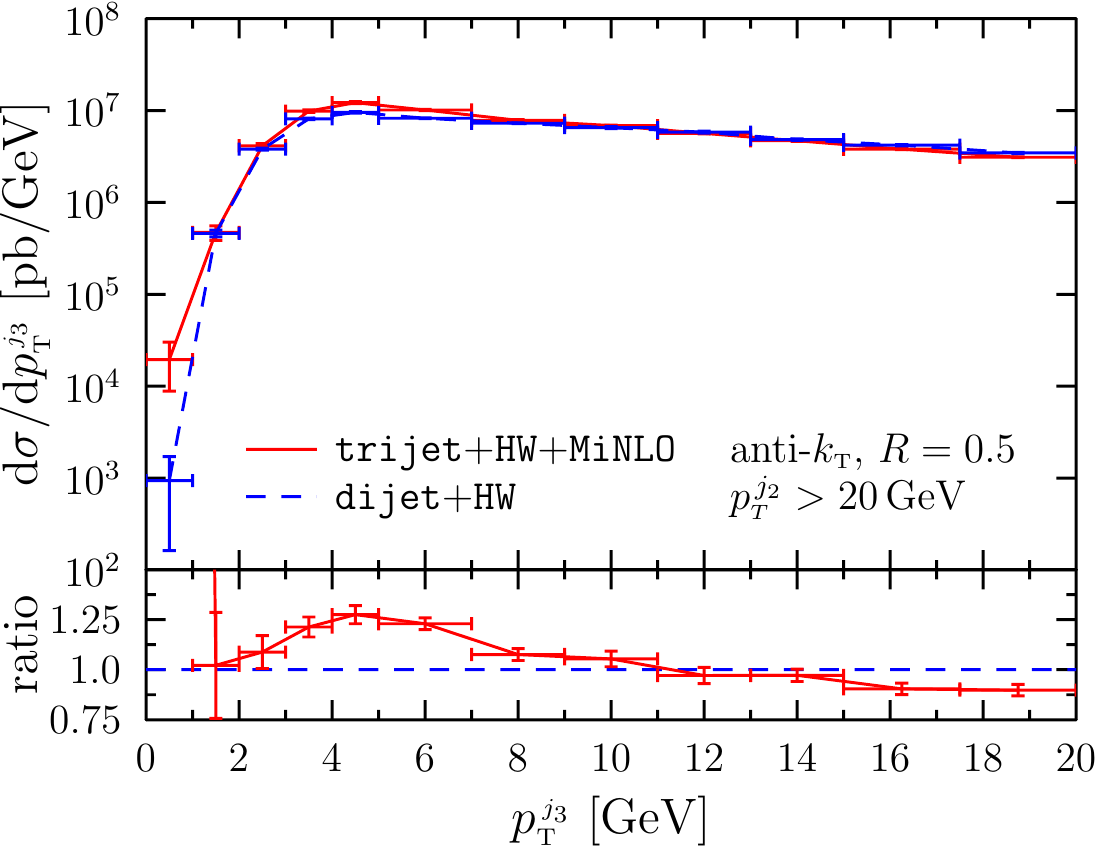,
   width=0.49\textwidth}
\epsfig{file=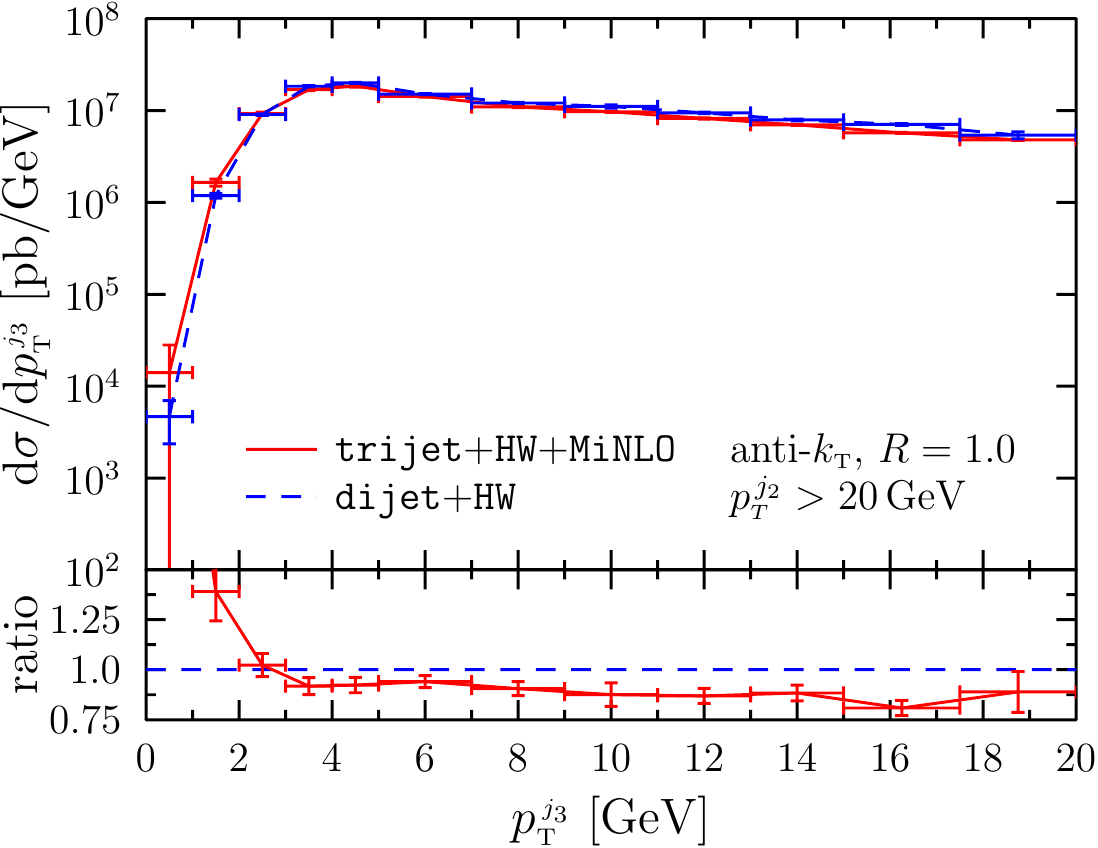,
   width=0.49\textwidth}
\caption{\label{fig:trijet-dijet-HW-j3} As in
  fig.~\protect{\ref{fig:trijet-dijet-PY8-j3}}, for \HERWIG{}}
\end{figure}

We begin by showing in fig.~\ref{fig:trijet-dijet-PY8-j3} the
transverse-momentum distribution of the third jet, $\pt^{j_3}$, in events
where $\pt^{j_2}> 20$~GeV.  The aim of the figure is to compare the Sudakov
effects affecting the production of the third jet, introduced by the
\POWHEG{} machinery, in the case of the \dijet{} generator, with those
introduced in the \trijet{} generator by the \MiNLO{} procedure. We see that,
in both cases, the small transverse-momentum region is properly regulated by
the Sudakov form factor. We remind the reader that, as far as the large
transverse-momentum region is concerned, the \trijet{} generator has NLO
accuracy for this observable, while the \dijet{} one is only LO accurate.
The slightly strange features at very low transverse momentum, that we
observe with the \PYTHIAEIGHT{} shower, are shared for the same observable by
the \PYTHIASIX{} result, while, in the \HERWIG{} case, we see a smoother
behaviour, as shown in figs.~\ref{fig:trijet-dijet-PY-j3}
and~\ref{fig:trijet-dijet-HW-j3}.

\begin{figure}
\epsfig{file=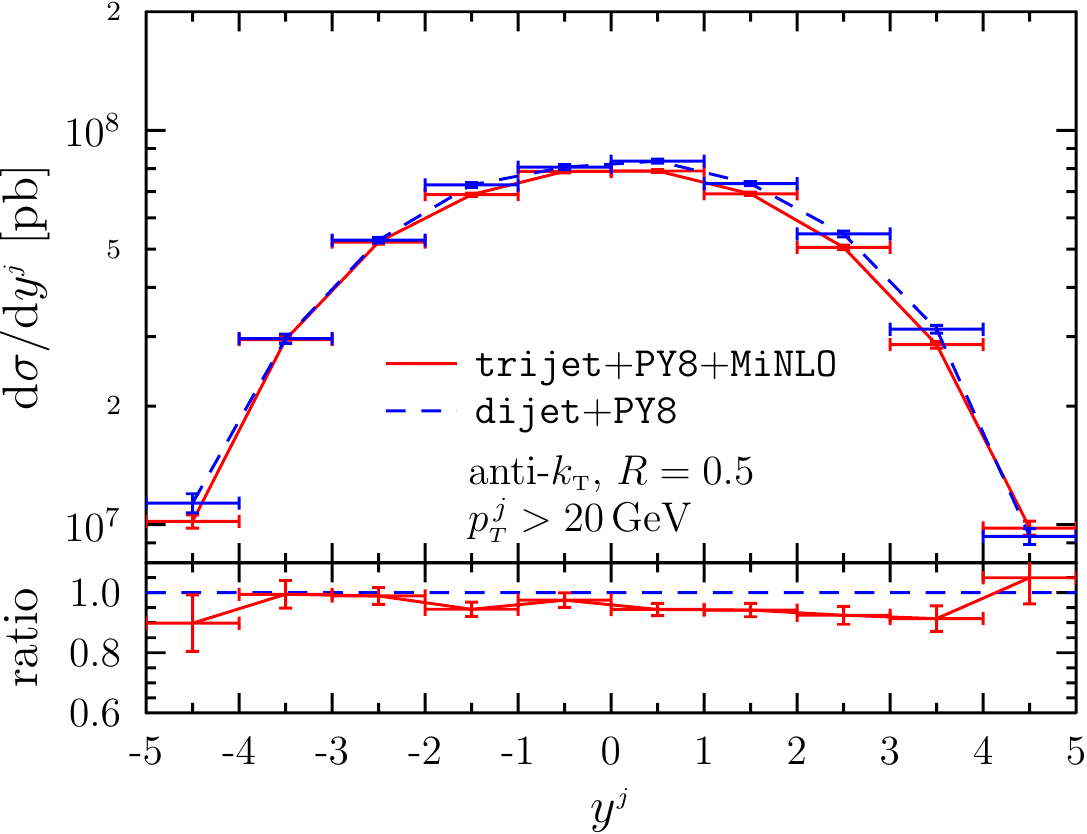,width=0.49\textwidth}
\epsfig{file=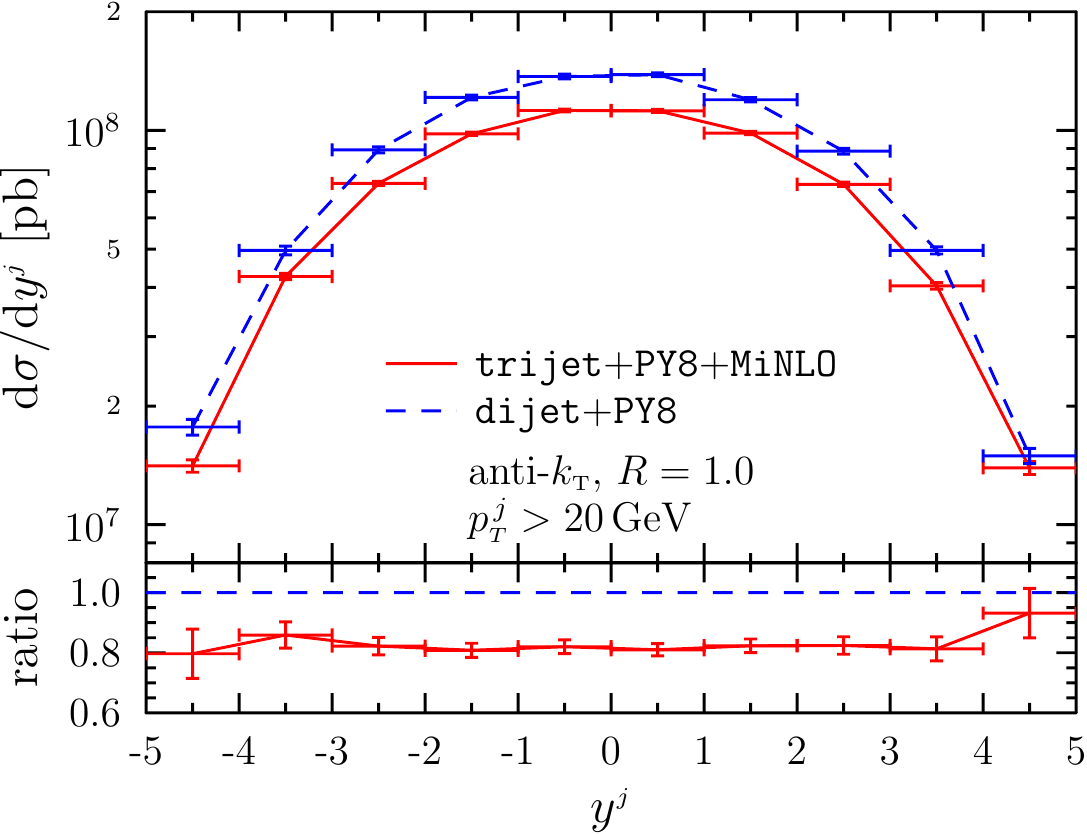,width=0.49\textwidth}
\caption{\label{fig:trijet-dijet-PY8_y} Comparison between the \PYTHIAEIGHT{}
  showered results for the rapidity distribution of an inclusive jet, for
  $R=0.5$~(left) and $R=1$~(right) for \trijet+\MINLO{} and \dijet{}. Jets
  are required to have transverse momentum larger than 20~GeV.}
\end{figure}
We now turn to the comparison of the \trijet{}+\MiNLO{} and \dijet{}
generators for inclusive-jet distributions.  We first compare the rapidity
distribution of an inclusive jet in fig.~\ref{fig:trijet-dijet-PY8_y}.  No
interesting differences are seen between the \trijet+\MiNLO{} and \dijet{}
curves, other than the obvious difference in normalization, due to the cut in
transverse momentum, that could also be evinced from the transverse-momentum
distribution.  We show this distribution only for the case of \PYTHIAEIGHT{}
shower, since we obtain similar results when using \HERWIG{} or \PYTHIA{}.

\begin{figure}
\epsfig{file=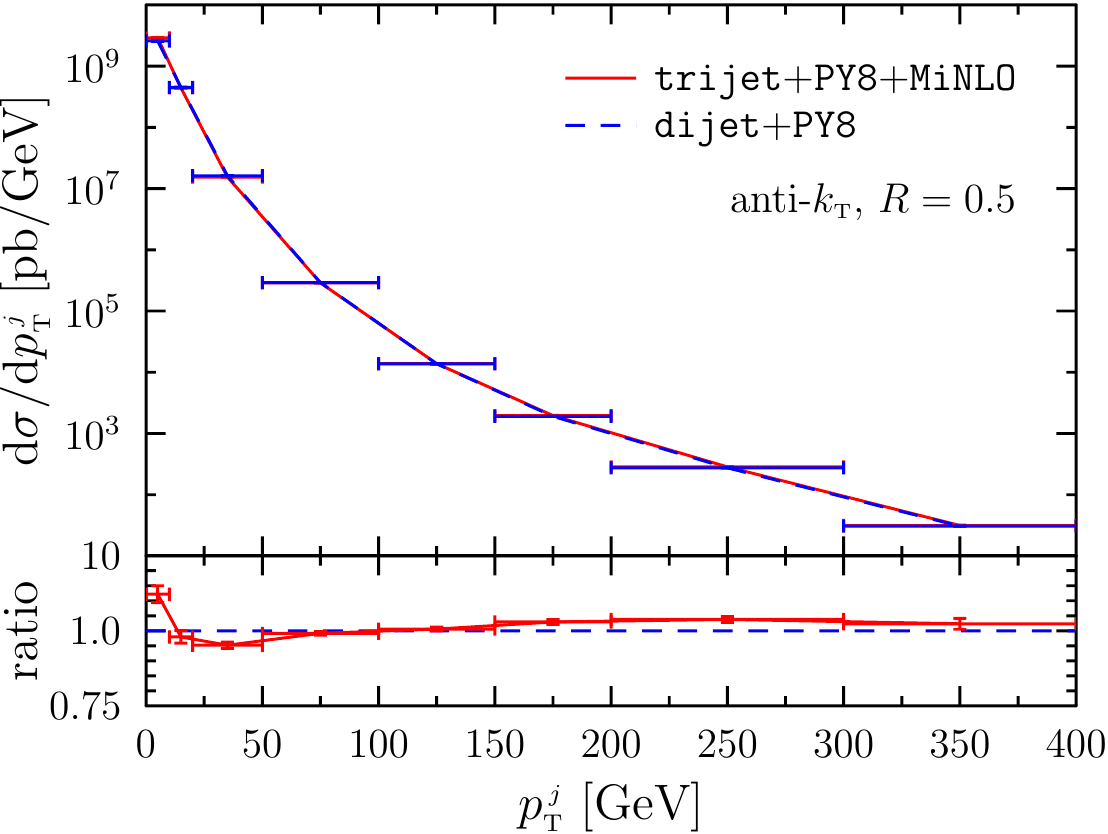,width=0.49\textwidth}
\epsfig{file=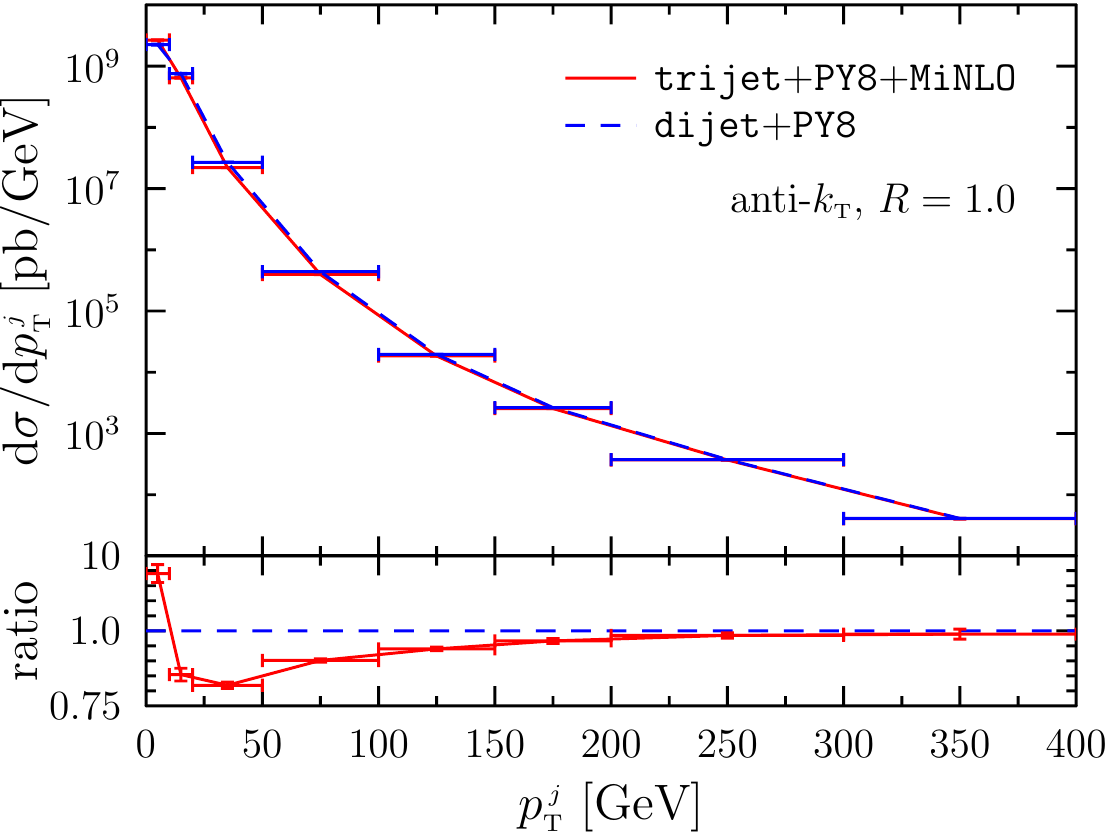,width=0.49\textwidth}
\caption{\label{fig:trijet-dijet-PY8} Comparison between the \PYTHIAEIGHT{}
  showered results for the transverse-momentum distribution of an inclusive
  jet, for $R=0.5$~(left) and $R=1$~(right) for \trijet+\MINLO{} and
  \dijet{}.}
\end{figure}
\begin{figure}
\epsfig{file=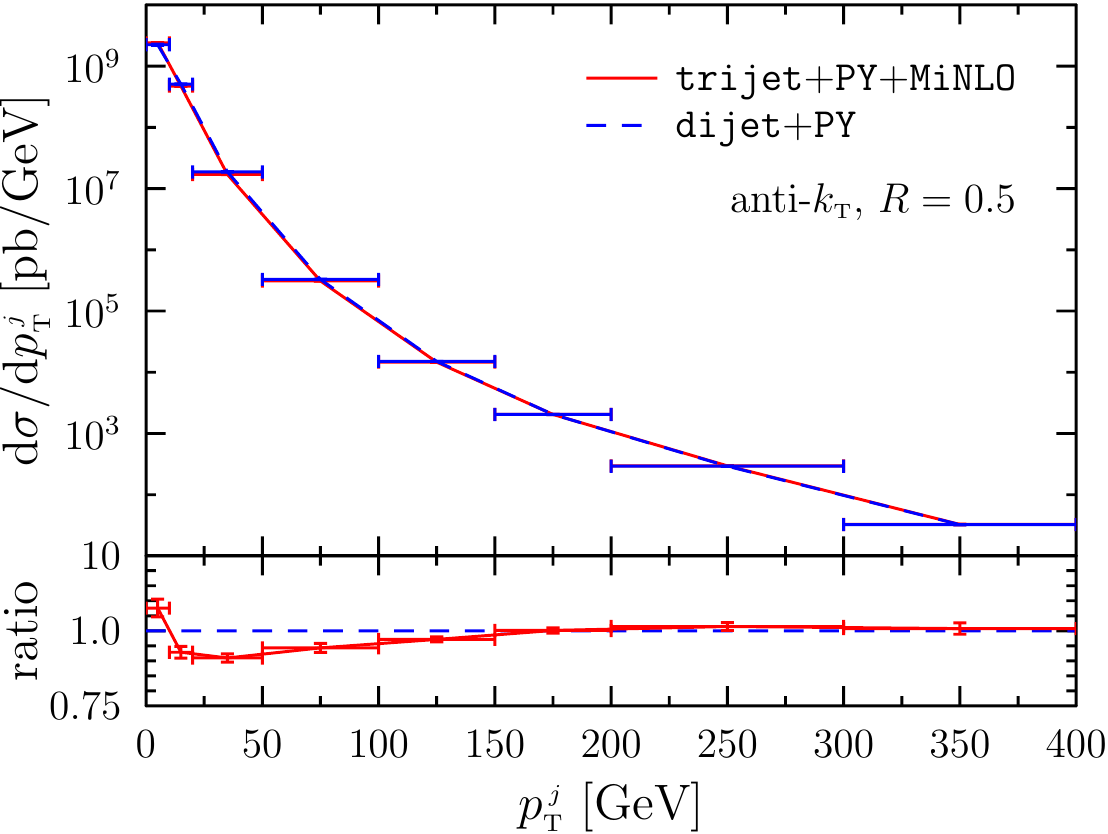,width=0.49\textwidth}
\epsfig{file=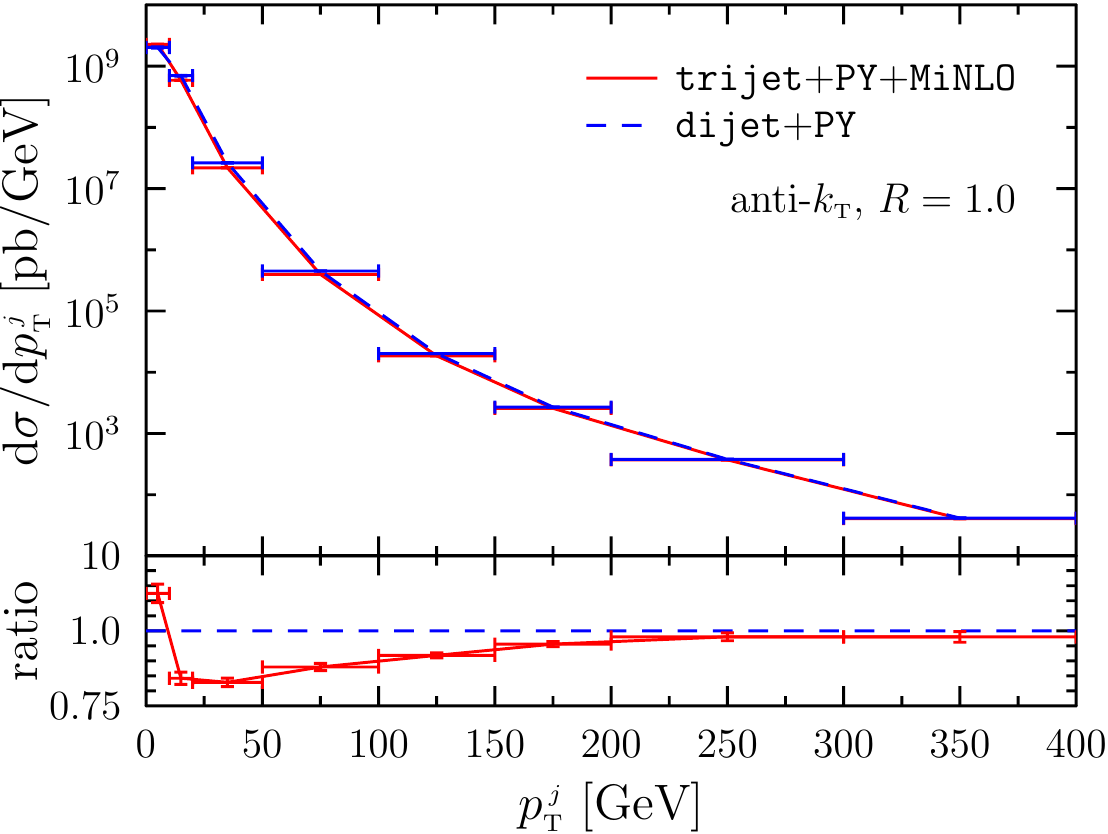,width=0.49\textwidth}
\caption{\label{fig:trijet-dijet-PY} Same as fig.~\ref{fig:trijet-dijet-PY8}
for \PYTHIA{}.}
\end{figure}
\begin{figure}
\epsfig{file=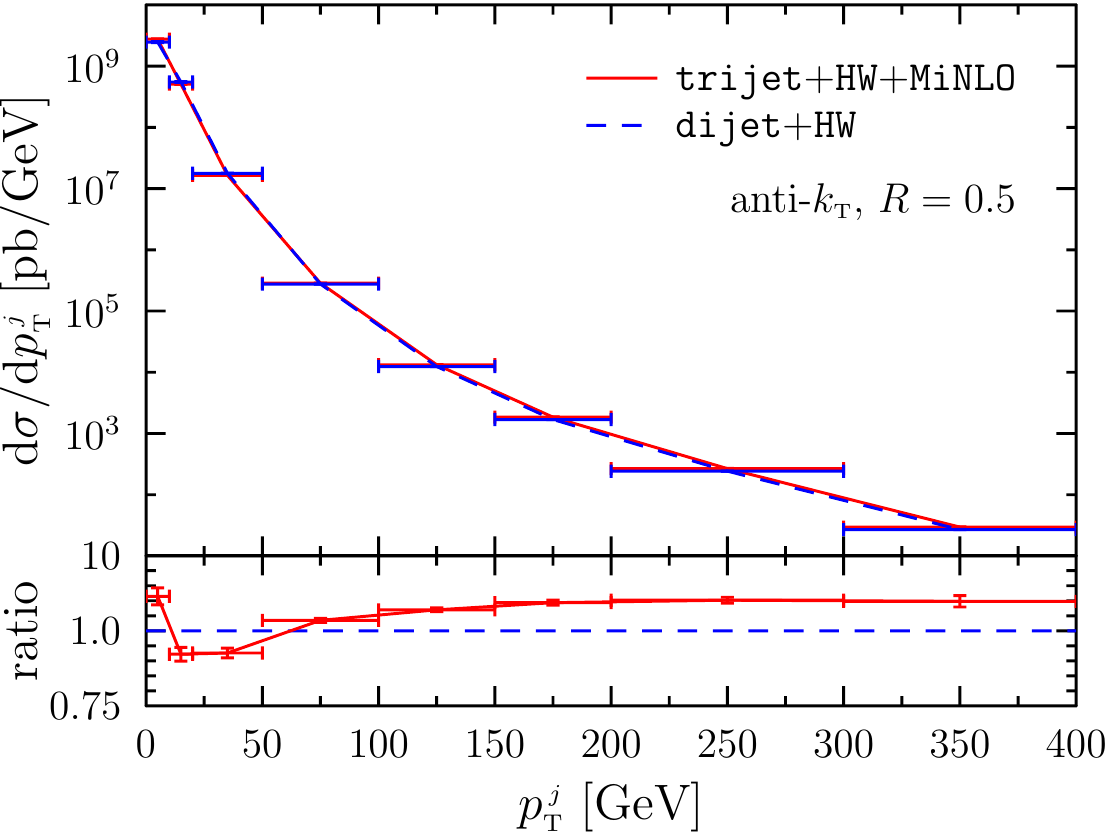,width=0.49\textwidth}
\epsfig{file=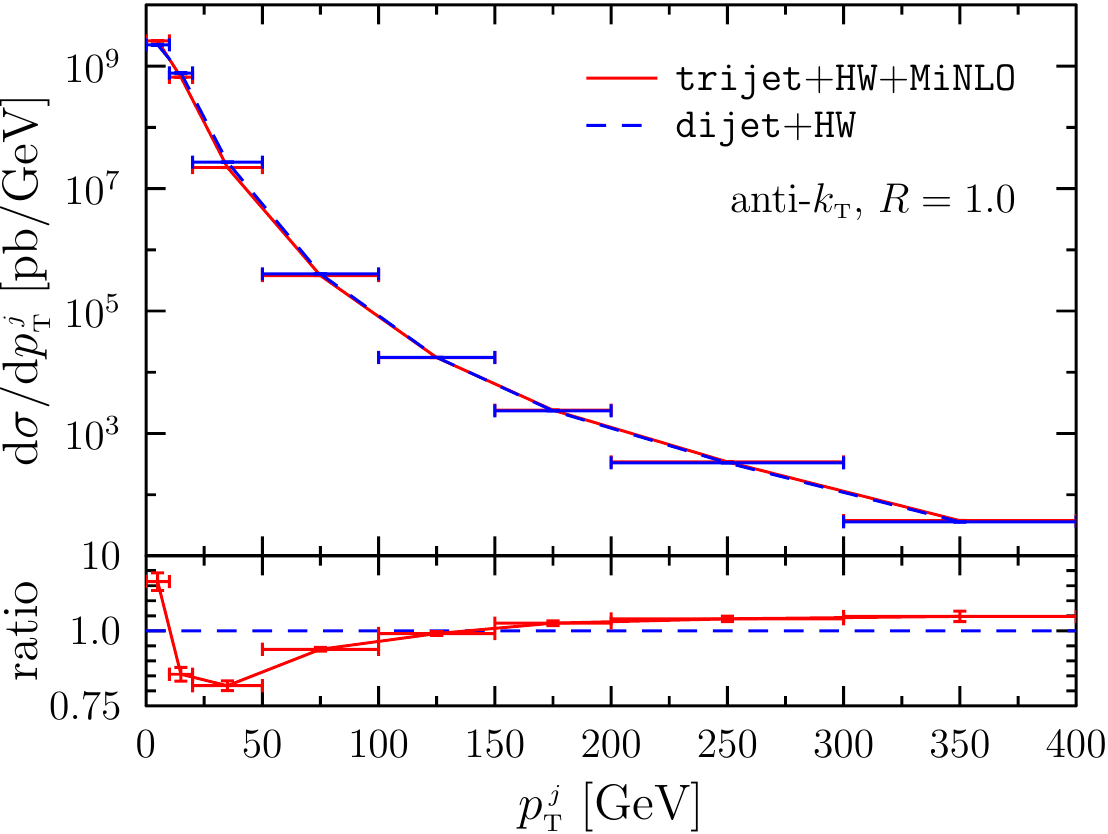,width=0.49\textwidth}
\caption{\label{fig:trijet-dijet-HW}  Same as fig.~\ref{fig:trijet-dijet-PY8}
for \HERWIG{}.}
\end{figure}
Another interesting distribution is the transverse-momentum of an inclusive
jet, displayed in figs.~\ref{fig:trijet-dijet-PY8}, \ref{fig:trijet-dijet-PY}
and~\ref{fig:trijet-dijet-HW} for \PYTHIAEIGHT{}, \PYTHIA{} and \HERWIG{}
respectively.  We notice the remarkable agreement between the
\trijet{}+\MiNLO{} and \dijet{} generators.

In conclusion, we find that the use of \MiNLO{} considerably improves the
\trijet{} generator also in the region where one jet becomes
unresolved. Observe that, while the \dijet{} generator is NLO accurate for
these distributions, the \trijet{}+\MiNLO{} generator is at most LO
accurate. Thus, we do not advocate the use of the \trijet{}+\MiNLO{}
generator for one or two jet inclusive distributions.  However, the fact that
also these regions are treated consistently gives us confidence that we
should not be afraid to make predictions for three-jet observables also in
the region where one jet is relatively close to being soft, or relatively
close to a collinear configuration. We thus recommend that our generator is
used with the \MiNLO{} feature turned on.

\subsection{Scale-variation bands}
\label{sec:scalevar}
\begin{figure}
\epsfig{file=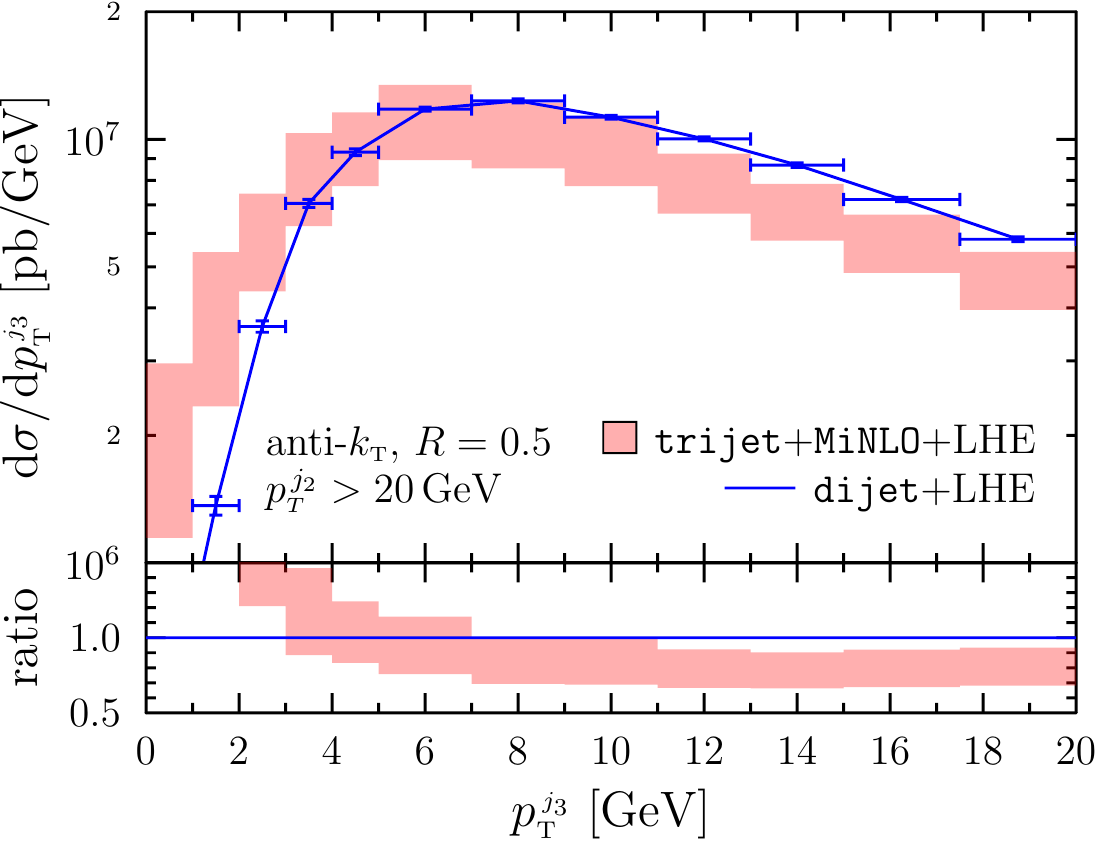,
   width=0.49\textwidth}
\epsfig{file=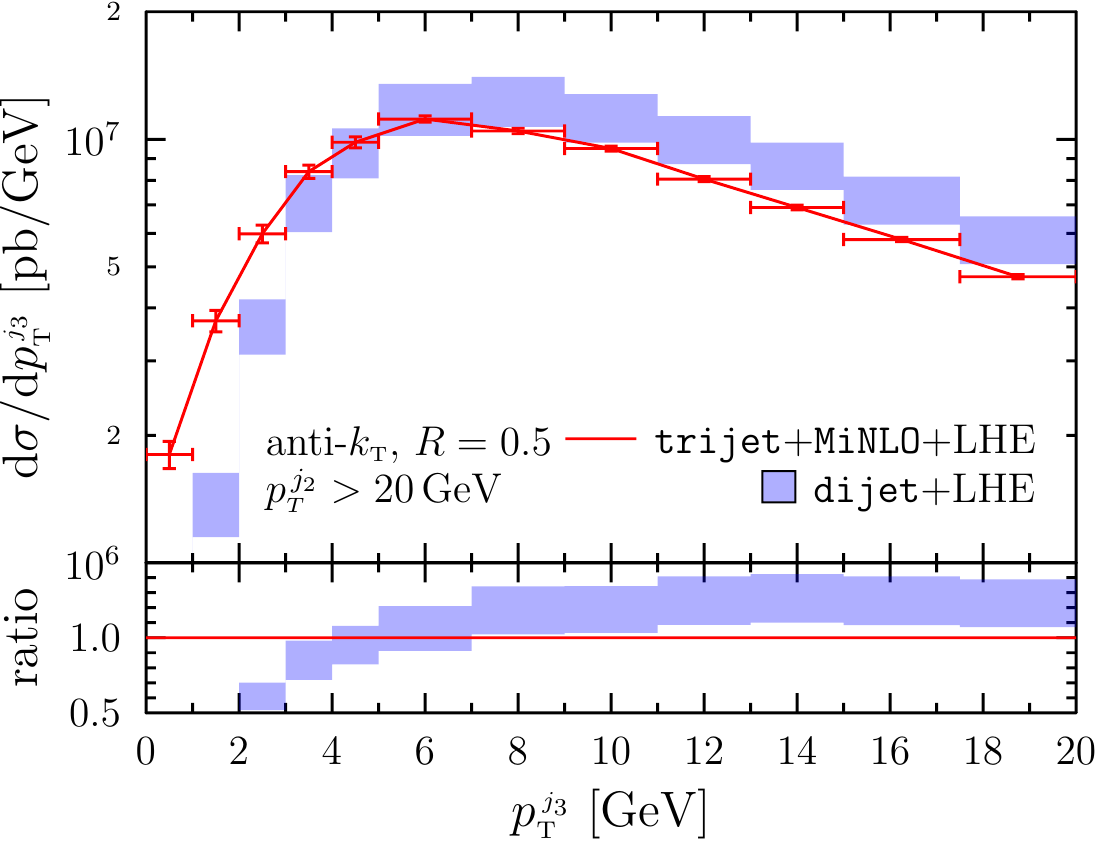,
   width=0.49\textwidth}
\caption{\label{fig:LH-scalevar-j3-ptzoom} Comparison between the LHE 7-point
  scale-variation band, for the transverse momentum of the third jet, with
  $R=0.5$. A minimum $\pt$ of 20~GeV is imposed on the second jet.  On the
  left plot, \trijet+\MINLO{} scale-variation band compared to the
  central-scale differential cross section for dijet production. On the
  right, \dijet{} scale-variation band compared to the central-scale cross
  section for trijet production.}
\end{figure}
\begin{figure}
\epsfig{file=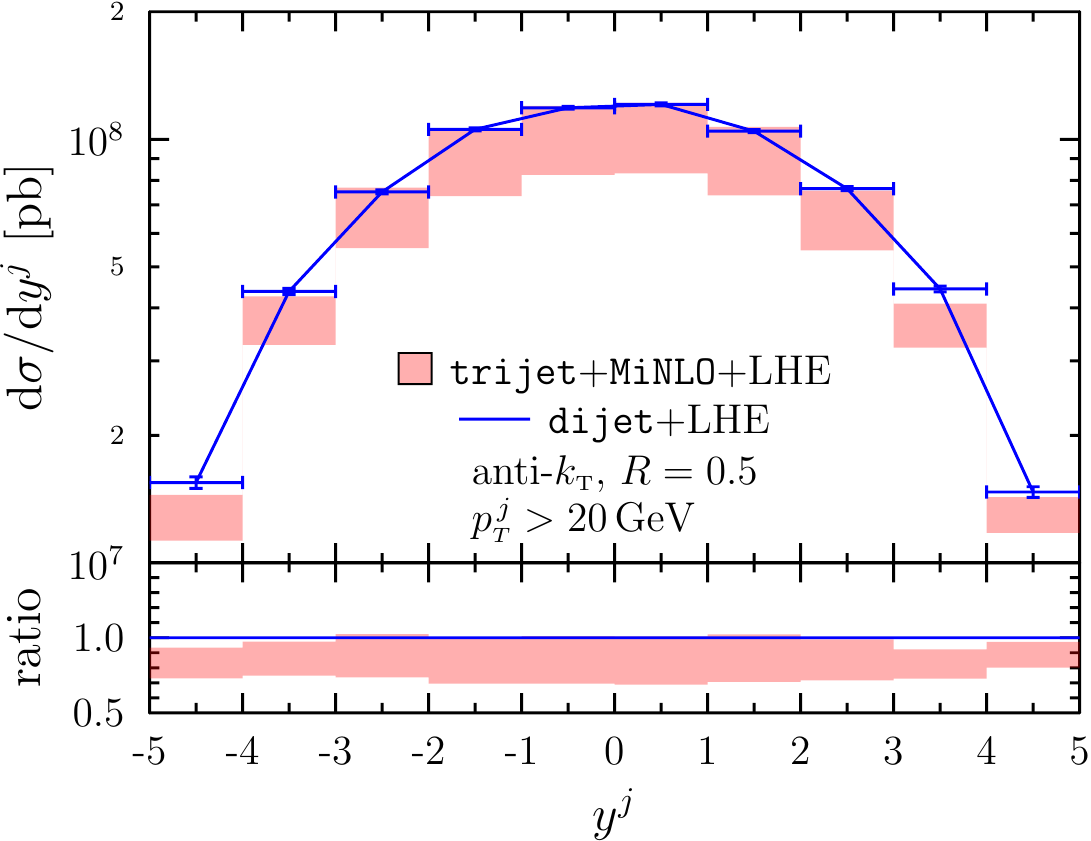,width=0.49\textwidth}
\epsfig{file=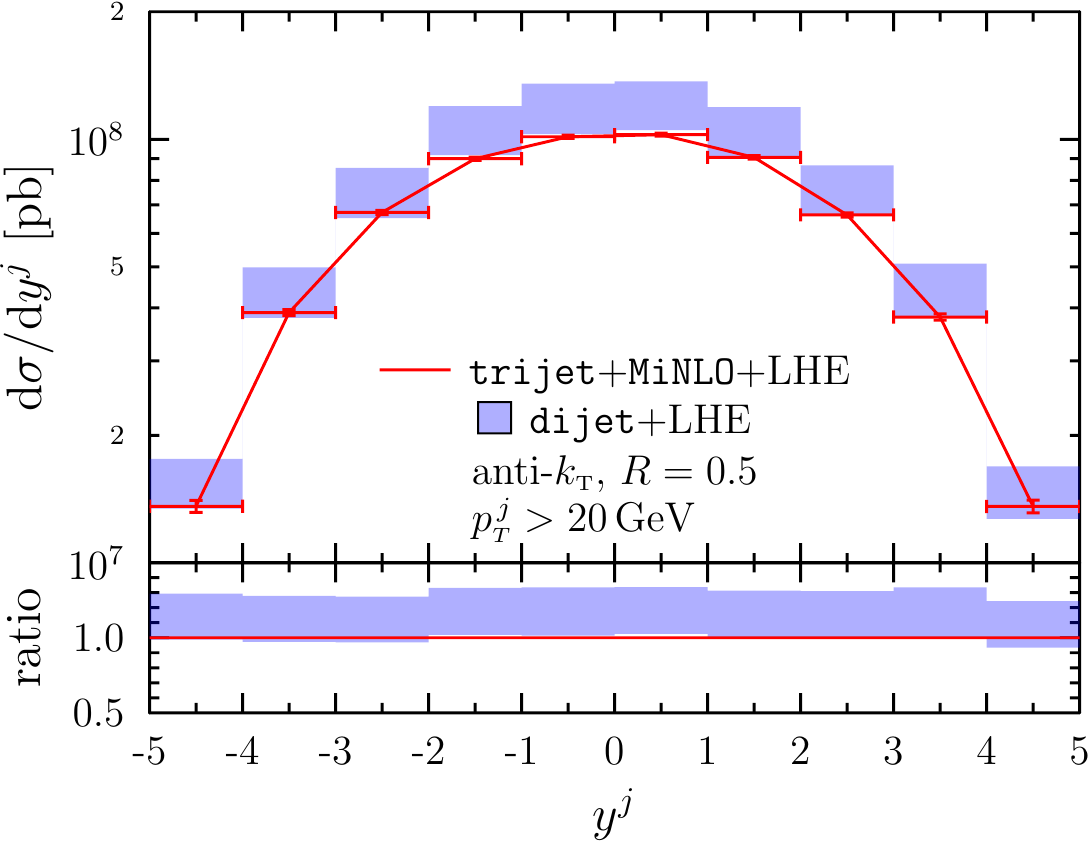,width=0.49\textwidth}
\caption{\label{fig:LH-scalevar-j-y} Same as
  fig.~\protect{\ref{fig:LH-scalevar-j3-ptzoom}} for the rapidity of the
  inclusive jet.  A minimum $\pt$ of 20~GeV is imposed on jets.}
\end{figure}
\begin{figure}
\epsfig{file=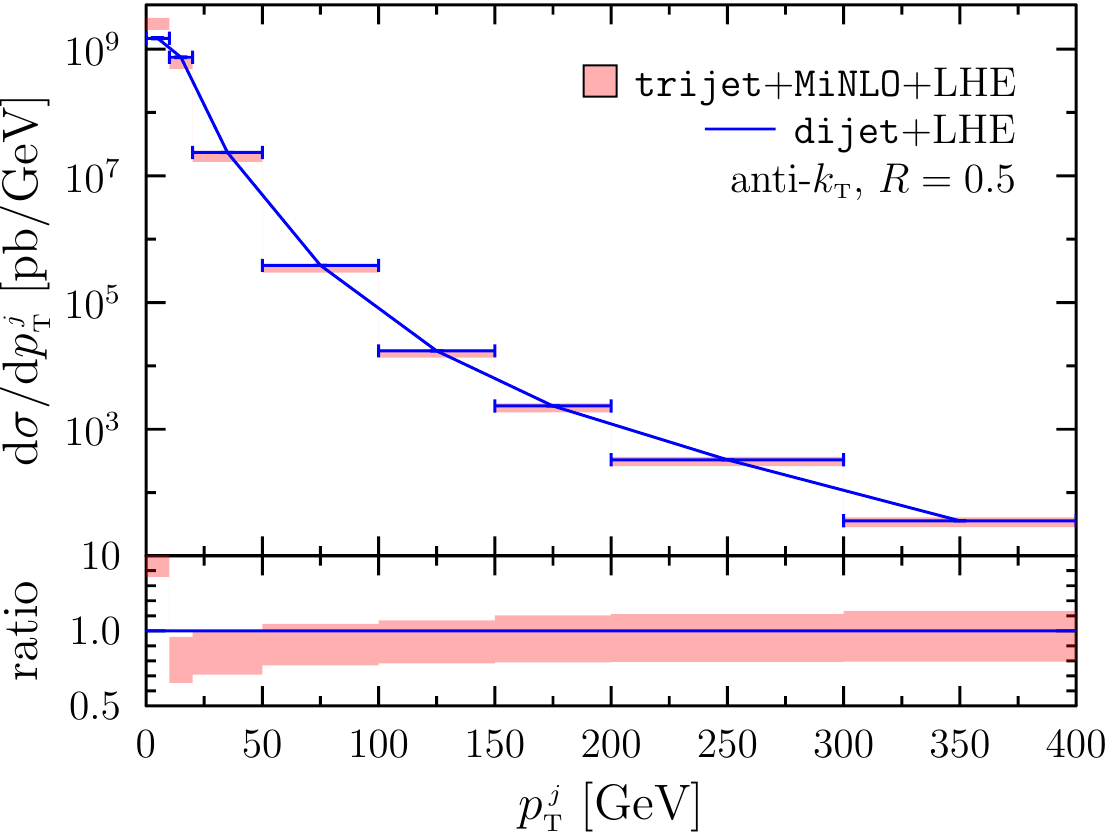,
   width=0.49\textwidth}
\epsfig{file=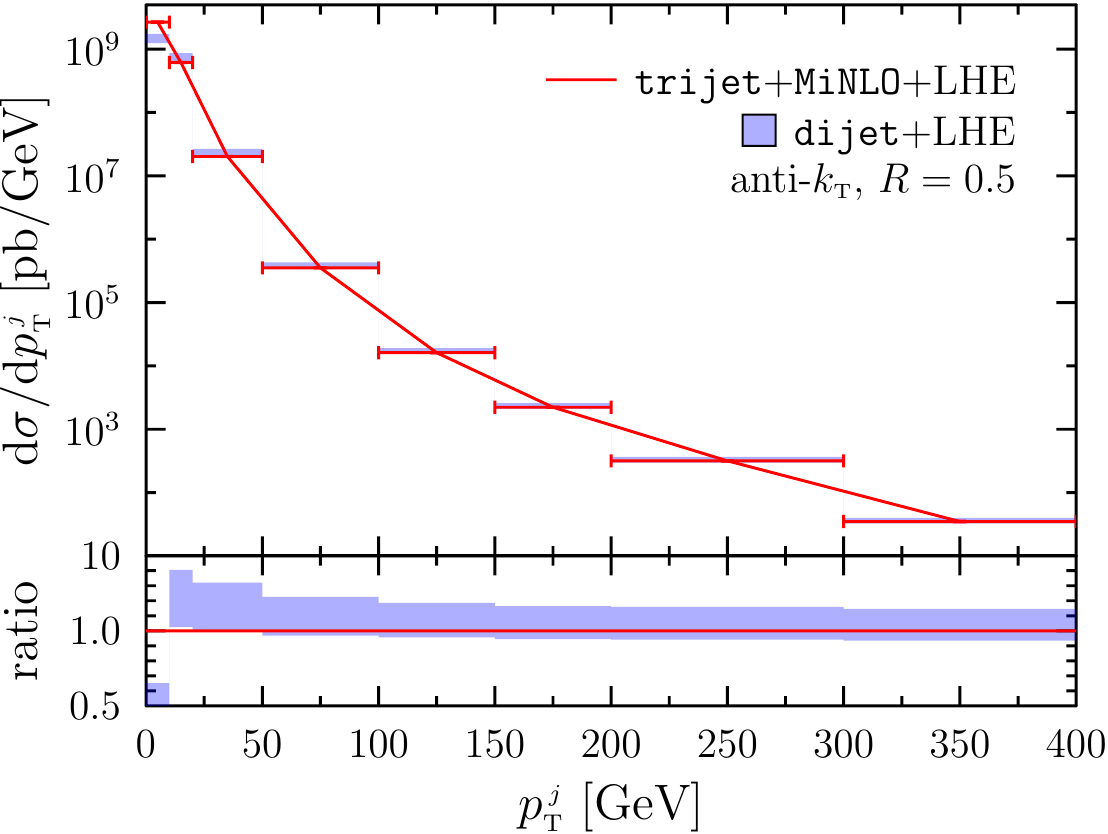,
   width=0.49\textwidth}
\caption{\label{fig:LH-scalevar-j-pt} Same as
  fig.~\protect{\ref{fig:LH-scalevar-j3-ptzoom}} for the transverse momentum
  of the inclusive jet.}
\end{figure}
In this section we show the scale-variation bands for some key distributions.
The purpose of this section is threesome: to show that scale uncertainties
can be easily computed, to give an idea of the uncertainty involved, and to
show that the remarkable agreement of the \trijet{}+\MiNLO{} and \dijet{}
results for inclusive quantities is not accidental.  A more thorough
uncertainty study will be carried out in a forecoming publication where
comparisons with available data will be considered.

We illustrate in figs.~\ref{fig:LH-scalevar-j3-ptzoom},
\ref{fig:LH-scalevar-j-y} and~\ref{fig:LH-scalevar-j-pt} the scale-variation
bands for the transverse-momentum distribution of the third hardest jet, and
the rapidity distribution and transverse momentum of the inclusive jet, for
\trijet{}+\MINLO{} at the LHE level. We compare the \trijet{}+\MiNLO{} scale
variation with the same scale variation in \dijet{} production.

The scale variation can be performed in a fast way, without regenerating the
event sample, using the reweighting tool in the \POWHEGBOXVtwo{}, that allows
for a very fast re-evaluation of the weight associated to each
event.\footnote{This feature is documented in the file {\tt
    README.LesHouchesReweighting} in the {\tt Docs} directory of
  \POWHEGBOXVtwo{}.}  The scale-variation band is obtained by taking the
envelope of the 7 differential cross sections computed by multiplying the
reference factorization and renormalization scales by the factors $\KFA$ and
$\KRA$, respectively, where
\begin{equation}
(\KRA,\KFA)=(0.5,0.5), (0.5,1), (1,0.5), (1,1), (2,1), (1,2), (2,2).
\end{equation}
In all the three distributions we are showing, we get a comparable scale-band
size in the \trijet{}+\MiNLO{} and \dijet{} results, of the order
of 20\%, and there is a very good degree of overlapping for the inclusive-jet
rapidity and transverse momentum.

\section{Conclusions}
\label{sec:conclusions}
In this paper, we have presented an implementation of a NLO  plus
parton-shower generator for three-jet production, built in the framework of
the \POWHEGBOXVtwo{}.

We have compared key kinematic distributions at different levels: NLO, Les
Houches event level, and after the shower performed by \PYTHIA,
\PYTHIAEIGHT{} and \HERWIGSIX.  We found very good agreement between the NLO
and the \PYTHIA{} and \PYTHIAEIGHT{} results, for variables that are
correctly described by a fixed-order calculation.  Slightly worse agreement
is found between the NLO and the \HERWIG-showered results.

We have also applied the recently-proposed \MiNLO{} procedure, for the scale
assignment in the NLO calculation, to our generator.  We have found that
\MiNLO{} considerably improves the \trijet{} generator also in the regions
where one jet becomes unresolved.  The fact that also these regions are
treated consistently gives us confidence that we can make predictions for
three-jet observables also in the region where one jet is relatively close to
being soft, or relatively close to a collinear configuration.

We have seen that the \trijet{}+\MiNLO{} and \dijet{} results display
remarkable consistency among each other. On the other hand, we have also
evidence that the kind of shower generator that is used for the final shower
has a non-negligible impact on the result, especially for relatively-small
jet cone sizes.

The code can be downloaded following the instructions in the \POWHEGBOX{} web
site \url{http://powhegbox.mib.infn.it}.

\section*{Acknowledgments}
We thank Simon Badger for helping us in the use of \NJET, and Gionata Luisoni
and Francesco Tramontano for their help in running the \GOSAM{} package.
A.K.~is grateful to Zolt\'an Nagy and Zolt\'an Tr\'ocs\'anyi for several
useful discussions and help with the \texttt{NLOJET++} code.  We kindly
acknowledge (A.K.~in particular) the financial support provided by the
LHCPhenoNet Training Network.  A.K.~acknowledges the Hungarian Scientific
Research Fund grant K-101482 and is thankful to the Aspen Center for Physics
for warm hospitality where part of this work was carried out.
The computations shown in this paper were partially performed on the HPC
facility of the University of Debrecen (NIIF Institute, Hungary).

\appendix

\section{Scale options in the \POWHEGBOX{}}
\label{app:btlscalereal}
In the \POWHEGBOX, the factorization and renormalization scales are usually
set as a function of the underlying-Born kinematics. Since the \POWHEGBOX{}
can also be used as a parton level, fixed-order generator (by setting {\tt
  testplots 1} in the {\tt powheg.input} file), it is convenient, at times,
to remove this restriction (for example, in order to compare the fixed order
NLO output to other codes). This is done as follows. If one sets the variable
{\tt btlscalereal 1} in the {\tt powheg.input} file, the internal flag {\tt
  flg\_btildepart} is used to distinguish the Born, virtual and
subtraction-term contributions from the real one.  When {\tt flg\_btildepart}
equals {\tt 'b'}, the program is computing the Born or virtual. When it is
set to {\tt 'r'} it is computing the real contribution. The user can then
modify the {\tt set\_fac\_ren\_scales} subroutine, so that, on the basis of
the value of {\tt flg\_btildepart}, the program uses the Born or real
kinematics to compute the scales.

Another ambiguity in the scale choice has to do with the computation of the
subtraction terms. It is acceptable to use for them the same scales used for
the real contributions. On the other hand, it is also acceptable to use for
them the scales of the corresponding underlying-Born configuration. In order
to implement also this option, one sets the variable {\tt btlscalect 1} in
the {\tt powheg.input} file. If this variable is set, the {\tt
  set\_fac\_ren\_scales} subroutine is called with {\tt flg\_btildepart}
equals to {\tt 'c'} when the subtraction terms are computed.

For the comparison with the results of ref.~\cite{Badger:2012pf}, we set
$\mur=\muf=\Ht/2$.  $\Ht$ is computed using the Born kinematics, for the
Born, the virtual and subtraction terms, and using the real-contribution
kinematics for the real terms. A code that implements this choice of scales
has the form:
\begin{verbatim}
      subroutine set_fac_ren_scales(muf,mur)
      ...
      include 'pwhg_kn.h'
      include 'pwhg_flg.h'
      ...
      if ((flg_btildepart.eq.'b').or.(flg_btildepart.eq.'c')) then
         pt1 = sqrt(kn_cmpborn(1,3)**2+kn_cmpborn(2,3)**2)
         pt2 = sqrt(kn_cmpborn(1,4)**2+kn_cmpborn(2,4)**2)
         pt3 = sqrt(kn_cmpborn(1,5)**2+kn_cmpborn(2,5)**2)
         Ht = pt1 + pt2 + pt3
      elseif ((flg_btildepart.eq.'r')) then
         pt1 = sqrt(kn_cmpreal(1,3)**2+kn_cmpreal(2,3)**2)
         pt2 = sqrt(kn_cmpreal(1,4)**2+kn_cmpreal(2,4)**2)
         pt3 = sqrt(kn_cmpreal(1,5)**2+kn_cmpreal(2,5)**2)
         pt4 = sqrt(kn_cmpreal(1,6)**2+kn_cmpreal(2,6)**2)
         Ht = pt1 + pt2 + pt3 + pt4
      endif
      muf=Ht/2
      mur=Ht/2
\end{verbatim}
where {\tt kn\_cmpborn} and {\tt kn\_cmpreal} are the arrays of the Born and
real center-of-mass momenta, defined in the \POWHEGBOX{} {\tt pwhg\_kn.h}
include file, {\tt flg\_btildepart} is declared in the {\tt pwhg\_flg.h}
file, and {\tt pt1} \ldots {\tt pt4} are local variables denoting the
transverse momenta of the final-state partons.


\providecommand{\href}[2]{#2}\begingroup\raggedright\endgroup

\end{document}